\documentclass[acmstoms,natbib=false]{acmart}

\usepackage{upgreek}
\usepackage{subcaption}
\usepackage{xcolor}
\definecolor{darkgreen}{rgb}{0.0, 0.5, 0.0}
\definecolor{darkred}{rgb}{0.5, 0.0, 0.0}
\usepackage{arydshln} 
\usepackage{rotating}  
\usepackage{pdflscape} 
\usepackage{placeins}
\usepackage{multirow}

\usepackage{xspace}
\usepackage[acronym]{glossaries}
\glsdisablehyper  
\newacronym[plural=CPUs, longplural=Central Processing Units]{cpu}{CPU}{Central Processing Unit}
\newacronym[plural=GPUs, longplural=Graphics Processing Units]{gpu}{GPU}{Graphics Processing Unit}
\newacronym[plural=ASICs, longplural=Application-Specific Integrated Circuit]{asic}{ASIC}{Application-Specific Integrated Circuit}
\newacronym{ram}{RAM}{Random Access Memory}
\newacronym{tlb}{TLB}{Translation Lookaside Buffer}
\newacronym{ai}{AI}{Aritificial Intelligence}
\newacronym{abi}{ABI}{Application Binary Interface}
\newacronym[plural=flops, longplural=Floating Point Operations]{flop}{FLOP}{Floating Point Operation}
\newacronym{hpc}{HPC}{High Performance Computing}
\newacronym{bsp}{BSP}{Bulk Synchronous Parallel}
\newacronym{mpi}{MPI}{Message Passing Interface}
\newacronym{jit}{JIT}{Just In Time}
\newacronym{cuda}{CUDA}{Compute Unified Device Architecture}
\newacronym{rdma}{RDMA}{Remote Direct Memory Access}
\newacronym{aot}{AOT}{Ahead of Time}
\newacronym[plural=ISAs, longplural=Instruction Set Architectures]{isa}{ISA}{Instruction Set Architecture}
\newacronym{raii}{RAII}{Resource Allocation is Initialisation}
\newacronym[plural=CCXs, longplural=Core Complexes]{ccx}{CCX}{Core Complex}
\newacronym[plural=CCDs, longplural=Core Complex Dies]{ccd}{CCD}{Core Complex Die}

\newacronym[plural=SVDs]{svd}{SVD}{Singular Value Decomposition}
\newacronym[plural=rSVDs]{rsvd}{rSVD}{Randomized Singular Value Decomposition}
\newacronym{cpqr}{CPQR}{Column Pivoted QR}
\newacronym{qr}{QR}{QR}
\newacronym{id}{ID}{Interpolative Decomposition}
\newacronym{aca}{ACA}{Adaptive Cross Approximation}
\newacronym{cur}{CUR}{Pseudo-Skeletal}
\newacronym{hodlr}{HODLR}{Hierarchical Off-Diagonal Low Rank}
\newacronym{hbs}{HBS/HSS}{Hierarchical Block Separable/Hierarchical Semi Separable}
\newacronym{blas_m2l}{BLAS-M2L}{BLAS Based Multipole to Local (M2L) Field Translation}
\newacronym{fft_m2l}{FFT-M2L}{FFT Based Multipole to Local (M2L) Field Translation}
\newacronym{h}{$\mathcal{H}$}{$\mathcal{H}$ Matrix}
\newacronym{htwo}{$\mathcal{H}^2$}{$\mathcal{H}^2$ Matrix}

\newacronym{fma}{FMA}{Fused Multiply and Add}
\newacronym{sisd}{SISD}{Single Instruction Single Data}
\newacronym{simd}{SIMD}{Single Instruction Multiple Data}
\newacronym{mimd}{MIMD}{Multiple Instruction Multiple Data}
\newacronym{simt}{SIMT}{Single Instruction Multiple Threads}

\newacronym{ilp}{ILP}{Instruction Level Parallelism}
\newacronym{tlp}{TLP}{Thread Level Parallelism}
\newacronym{dlp}{DLP}{Data Level Parallelism}
\newacronym{clp}{CLP}{Core Level Parallelism}

\newacronym{dram}{DRAM}{Dynamic Random Access Memory}
\newacronym{neon}{Neon}{Arm SIMD Extensions}
\newacronym{sse}{SSE}{Streaming SIMD Extensions}
\newacronym{avx}{AVX}{Advanced Vector Extensions/Gesher New Instructions}
\newacronym{avx2}{AVX2}{Advanced Vector Extensions/Haswell New Instructions}
\newacronym{avx512}{avx-512}{advanced vector extensions 512 bit}

\newacronym{blas}{BLAS}{Basic Linear Algebra Subprograms}
\newacronym{l2l}{L2L}{Local to Local}
\newacronym{l2p}{L2P}{Local to Particle}
\newacronym{m2l}{M2L}{Multipole to Local}
\newacronym{m2p}{M2L}{Multipole to Particle}
\newacronym{m2m}{M2M}{Multipole to Multipole}
\newacronym{p2p}{P2P}{Particle to Particle}
\newacronym{p2m}{P2M}{Particle to Multipole}

\newacronym[plural=kiFMMs, longplural=Kernel Independent Fast Multipole Methods]{kifmm}{kiFMM}{Kernel Independent Fast Multipole Method}
\newacronym[plural=bbFMMs, longplural=Black Box Fast Multipole Methods]{bbfmm}{bbFMM}{Black Box Fast Multipole Method}
\newacronym[plural=FMMs, longplural=Fast Multipole Methods]{fmm}{FMM}{Fast Multipole Method}
\newacronym[plural=PDEs]{pde}{PDE}{Partial Differential Equation}
\newacronym{mfs}{MFS}{Method of Fundamental Solutions}
\newacronym[plural=BIEs, longplural=Boundary Integral Equations]{bie}{BIE}{Boundary Integral Equation}
\newacronym[plural=BEMs, longplural=Boundary Element Methods]{bem}{BEM}{Boundary Element Method}
\newacronym[plural=FFTs, longplural=Fast Fourier Transforms]{fft}{FFT}{Fast Fourier Transform}
\newacronym[plural=DFTs, longplural=Discrete Fourier Transforms]{dft}{DFT}{Discrete Fourier Transform}
\newacronym{epcc}{EPCC}{Edinburgh Parallel Computing Centre}
\newacronym[plural=LETs, longplural=Locally Essential Trees]{let}{LET}{Locally Essential Tree}

\newcommand{\blas}{\acrshort{blas}\xspace}
\newcommand{\fftfull}{\acrfull{fft}\xspace}

\newcommand{\blasmtl}{\acrshort{blas_m2l}\xspace}

\newcommand{\fftmtl}{\acrshort{fft_m2l}\xspace}

\newcommand{\fma}{\acrshort{fma}\xspace}
\newcommand{\fmafull}{\acrfull{fma}\xspace}

\newcommand{\fmm}{\acrshort{fmm}\xspace}
\newcommand{\mfs}{\acrshort{mfs}\xspace}
\newcommand{\mfsfull}{\acrfull{mfs}\xspace}
\newcommand{\svd}{\acrshort{svd}\xspace}
\newcommand{\simd}{\acrshort{simd}\xspace}
\newcommand{\rsvd}{\acrshort{rsvd}\xspace}
\newcommand{\rsvdfull}{\acrfull{rsvd}\xspace}

\newcommand{\fmmfull}{\acrfull{fmm}\xspace}
\newcommand{\kifmm}{\acrshort{kifmm}\xspace}

\newcommand{\kifmmfull}{\acrfull{kifmm}\xspace}
\newcommand{\bbfmm}{\acrshort{bbfmm}\xspace}
\newcommand{\bbfmmfull}{\acrfull{bbfmm}\xspace}
\newcommand{\fmms}{\acrshortpl{fmm}\xspace}
\newcommand{\mtl}{\acrshort{m2l}\xspace}
\newcommand{\mtlfull}{\acrfull{m2l}\xspace}
\newcommand{\fft}{\acrshort{fft}\xspace}
\newcommand{\dft}{\acrshort{dft}\xspace}
\newcommand{\dftfull}{\acrfull{dft}\xspace}
\newcommand{\mtm}{\acrshort{m2m}\xspace}

\newcommand{\ltl}{\acrshort{l2l}\xspace}

\newcommand{\ltp}{\acrshort{l2p}\xspace}

\newcommand{\ptp}{\acrshort{p2p}\xspace}
\newcommand{\ptpfull}{\acrfull{p2p}\xspace}

\newcommand{\ptmfull}{\acrfull{p2m}\xspace}
\newcommand{\cpu}{\acrshort{cpu}\xspace}
\newcommand{\cpus}{\acrshortpl{cpu}\xspace}
\newcommand{\gpu}{\acrshort{gpu}\xspace}
\newcommand{\gpus}{\acrshortpl{gpu}\xspace}

\newcommand{\flops}{\acrshortpl{flop}\xspace}

\def\real{ \mathbb{R}}

\def\rthree{\real^3}

\def \xbf{\mathbf{x}}
\def \ybf{\mathbf{y}}

\def \ybfj{\mathbf{y}_j}
\def \xbfi{\mathbf{x}_i}

\usepackage{eucal}
\usepackage{bm}
\def\bigO#1{\mathcal{O}\left(#1\right)}
\newcommand{\mathbsf}[1]{\mathsf{\mathbf{#1}}}

\newcommand{\boldcolor}[2]{\textbf{\textcolor{#1}{#2}}}

\AtBeginDocument{%
  }

\setcopyright{acmlicensed}
\copyrightyear{2025}
\acmYear{2025}
\acmDOI{XXXXXXX.XXXXXXX}

\begin{document}

\title[M2L Operators for Kernel-Independent FMM]{M2L Translation Operators for Kernel Independent Fast Multipole Methods on Modern Architectures}

\author{Srinath Kailasa}
\email{sk937@cam.ac.uk}
\orcid{0000-0001-9734-8318}
\authornotemark[1]
\affiliation{%
  \institution{University of Cambridge}
  \city{Cambridge}
  \state{England}
  \country{UK}
}

\author{Timo Betcke}
\email{t.betcke@ucl.ac.uk}
\orcid{0000-0002-3323-2110}
\authornotemark[2]
\affiliation{%
  \institution{University College London}
  \city{London}
  \state{England}
  \country{UK}
}

\author{Sarah El Kazdadi}
\email{sarahelkazdadi@gmail.com}
\orcid{0000-0002-5657-0710}
\authornotemark[2]

\renewcommand{\shortauthors}{Kailasa et al.}

\begin{abstract}
Hardware trends favor algorithm designs that maximize data reuse per FLOP. We develop and benchmark high-performance Multipole-to-Local (M2L) translation operators for the kernel-independent Fast Multipole Method (kiFMM), a widely adopted FMM variant that supports a broad class of kernels and has been favored by recent implementations for its simple specification. Naively implemented, M2L is bandwidth-limited and therefore a key bottleneck in the FMM. State-of-the-art FFT-based M2L implementations, though elegant and with a fast setup time, suffer from low operational intensity and require architecture-specific optimizations. We demonstrate that a BLAS-based M2L, combined with randomized low-rank compression, achieves competitive performance with greater portability and a simpler implementation leveraging existing BLAS infrastructure, at the cost of higher setup times—especially for high-accuracy settings in double precision. Our Rust-based implementation enables seamless switching between strategies for fair benchmarking. Results on CPUs show that FFT-based M2L is favorable in low-accuracy settings or dynamic particle simulations, while BLAS-based M2L is favored for high-accuracy settings for static particle distributions, where its higher setup costs are amortized in many practical applications of the \fmm.
\end{abstract}



\received{20 February 2007}
\received[revised]{12 March 2009}
\received[accepted]{5 June 2009}

\maketitle

\section{Introduction}

The \fmmfull accelerates the evaluation of $N$-body sums of the form

\begin{equation}
  \phi_i = \sum_{j=1}^N K(\xbfi, \ybfj) q_j, \quad i = 1, \dots, M,
  \label{eq:sec:introduction:potential}
\end{equation}

where $\{\xbfi\}_{i=1}^M \subset \mathbb{R}^d$ are target points, $\{\ybfj\}_{j=1}^N \subset \mathbb{R}^d$ are source points, $q_j$ are source strengths, and $K : \mathbb{R}^d \times \mathbb{R}^d \to \mathbb{R}$ is an interaction kernel, where $d$ is the spatial dimension. The \fmm applies to asymptotically smooth, non-oscillatory kernels. A canonical example is the Laplace kernel,

\begin{equation}
  K(\xbf, \ybf) = \begin{cases}
\frac{1}{2 \pi} \log(\frac{1}{\|\xbf-\ybf\|}),  \> \> (d=2) \\
\frac{1}{4 \pi \|\xbf-\ybf\|}, \> \> (d=3)
  \end{cases}
  \label{eq:sec:introduction:laplace_kernel}
\end{equation}

which arises in electrostatics and gravitational interactions, and will be the focus of our benchmarks in this paper. For asymptotically smooth kernels, the \fmm reduces the naive $O(MN)$ to $O(P(N+M))$, where $P \ll N, M$ is the \textit{expansion order} controlling the approximation accuracy.

Equation \eqref{eq:sec:introduction:potential} can be interpreted as a dense matrix-vector product, where the matrix exhibits \textit{low rank} structure in its off-diagonal blocks. Such matrices occur frequently in computational science, extending the applicability of the \fmm from $N$-body simulations. For example, \fmms are widely used for accelerating dense matrix-vector products that arise in the iterative solution of boundary integral equations via the \acrfull{bem}, with applications from geophysics \cite{fujiwara2000fast}, to  fluid dynamics \cite{rahimian2010petascale}. Similar matrices appear in other fields, for example in data science applications such as in Kalman filtering or Gaussian processes \cite{li2014kalman, ambikasaran2014fast}.

The \fmm achieves its acceleration through a hierarchical decomposition of the problem domain, recursive algorithmic structure, and the construction of compressed representations for the \textit{far field}, corresponding to interactions between \textit{well-separated} clusters of source and target points. This allows the expensive direct evaluation of~\eqref{eq:sec:introduction:potential} to be limited to a \textit{near field} neighborhood of each cluster of target points. For a target cluster, the potential is split as

\begin{equation}
  \phi_i = \sum_{\ybfj \in \text{Near}(\xbfi)} K(\xbfi, \ybfj) q_j + \sum_{\ybfj \in \text{Far}(\xbfi)} K(\xbfi, \ybfj)q_j
  \label{eq:sec:introduction:near_far_split}
\end{equation}

where the first term represents direct, near-field interactions, and the second term represents approximated far-field interactions.

The key motivation for the development of the \fmm was the cost of the direct evaluation of the kernel, and hence performance was achieved by limiting the near-field interactions for each target cluster. However, in modern architectures the near-field interactions are highly suited for parallel implementation using either \acrshort{simd} or \acrshort{simt} programming paradigms, and it is the efficient handling of the far-field interactions that controls performance.

In particular, the evaluation of far-field approximations introduces memory bandwidth bottlenecks and non-contiguous memory access patterns, the impact of which relative to floating-point operations is significantly more important in terms of runtime performance on contemporary hardware \cite{dongarra2017extreme}. Consequently, optimizing the \mtlfull operation, which is used to summarize the far-field potential at each target cluster, has become critical for achieving high performance in \fmm implementations.

In this paper, we focus on \mtl as implemented for the \kifmmfull \cite{Ying2004}. This variant has been preferred in recent high-performance software implementations due to its straightforward specification in terms of matrix-product operations and its compatible with a wide variety of kernel functions of linear second order elliptic \acrshortpl{pde} \cite{Malhotra2015,wang2021exafmm}. Upon discretization, the matrices corresponding to the \mtl operator, which we call \mtl-matrices are known to be of \textit{low numerical rank}, and are therefore amenable to numerical compression. We refer to the \mtl implemented as a direct matrix product with low-rank compression as the \blasmtl approach.

Recent implementations of the \kifmm \cite{Malhotra2015, wang2021exafmm}, have relied on the \fftfull to accelerate the \mtl operation as it can be formulated as a convolution. We call this approach the \fftmtl, its algorithmic and \cpu-implementation details are summarized in Appendix \ref{appendix:fft_m2l}. The \fftmtl approach achieves high performance, however convolution type operations result in bandwidth limited element-wise products. Handling this requires careful memory layout optimizations and explicit \acrshort{simd} programming for each \acrfull{isa} in order to achieve acceptable \cpu performance, and efficient \gpu implementations remain unresolved. Despite this, \fftmtl was favored over \blasmtl in the original presentation of the \kifmm \cite{Ying2004} for three-dimensional problems. Though \blasmtl is algorithmically simpler—arising naturally from the \kifmm specification—its performance strongly depends on the low-rank compressibility of the \mtl matrices, making it more effective in two dimensions than in three. \fftmtl was later heavily optimized for x86 \cpus in PVFMM \cite{Malhotra2015}, with the same optimizations adopted by ExaFMM \cite{wang2021exafmm}. However, as \acrshort{blas} operations are increasingly optimized at the software and hardware levels \cite{VanZee2015, cabrera2021toward, gazzoni2024pqc}, we reconsider whether this operation can be formulated in terms of \blas operations to exploit these developments.

BLAS-based approaches for such problems were previously explored in the context of the \bbfmmfull, an alternative kernel-independent FMM that uses a Chebyshev basis for field representation \cite{Fong2009,messner2012optimized}. Our work is principally based on that of Fong et al, who introduced methods to reduce the cost of low-rank compression for \mtl matrices using the \svd for non-oscillatory kernels. Messner et al \cite{messner2012optimized} extended this with optimizations to lower further reduce the cost and introduced blocking strategies for CPU implementations to increase operational intensity of the matrix-matrix products for the \mtl. These strategies form the basis of their ScalFMM and TBFMM software \cite{bramas2020tbfmm,blanchard2015scalfmm,agullo2014task}.

Takahashi et al \cite{takahashi2012optimizing} developed an early \gpu implementation of the \blasmtl approach, implementing \gpu-specific blocking schemes. Their approach yielded promising results in single precision, but was constrained by the limited \gpu memory and high data transfer costs of that time, as well as the lack of efficient batched-\acrshort{blas} libraries such as those now available \cite{NVIDIA_cuBLAS}.

Our work builds upon ideas presented for the \bbfmm, where we show that the batch processing schemes described by Messner et al \cite{messner2012optimized} can be further simplified and organized to maximize operational intensity \footnote{We define operational intensity as the number of operations performed per unit of data retrieved from \textit{main memory} and subsequently filtered through a processor's cache hierarchy \cite{williams2009roofline}. Thus operational intensity is characterised by the traffic between a processor's cache and main memory, rather than \textit{within} the cache hierarchy of a processor - which is usually captured by a related concept called \textit{arithmetic intensity}. We prefer operational intensity in comparison to arithmetic intensity as the data we are concerned with is often too large to fit into a processor's cache.}. Furthermore, we show that the \svd-based compression scheme presented in \cite{Fong2009} can be accelerated with randomized methods \cite{halko2011finding} to speedup the required precomputations.

With this, we investigate whether the \blasmtl can be competitive with respect to the state-of-the-art \fftmtl for the \kifmm in three dimensions on modern \cpus. In order to provide a fair environment for comparison we have developed our own \kifmm implementation using Rust, a modern systems programming language emphasizing performance and safety. Our implementation, based on Rust's \texttt{trait} system enables the selection of different implementations of the \mtl operation. We have made sure to optimize the implementation to achieve similar levels of performance as PVFMM \cite{Malhotra2015} in order to have a good baseline for benchmarking.

We find that our \blasmtl approach is competitive with the \fftmtl approach of recent software, and maintains an edge for high-accuracy evaluations in double precision due to its high operational intensity. The trade-off with respect to the \fftmtl approach is a relatively longer precomputation time especially for high-accuracy simulations, however this is amortized in many practical applications such as in \acrshort{bem} solvers, where other parts of the setup far exceed the \fmm setup, or for problems in which precomputations can be reused - such as in the evaluation of the \fmm over multiple input source density vectors where the source/target positions are static. Furthermore, our \blasmtl benefits from a simple, and portable, algorithmic specification and is easy to translate in to a \gpu implementation based on batched-BLAS. The data organization required in our approach is also simpler in comparison to previous \blasmtl approaches, and we reduce the number of \acrshort{blas} calls to a minimum such that they maximize operational intensity. Our open-source software\footnote{Available on GitHub: \href{https://github.com/bempp/kifmm/}{https://github.com/bempp/kifmm/}}, allows us to perform a direct comparison with \fftmtl for the \kifmm while keeping constant the remainder of the software machinery. To the best of our knowledge there has been no comparison between these approaches, though similar work has been conducted for the analytical \fmm based on multipole expansions \cite{coulaud2008high, coulaud2010high}.

Our main contributions are:

\begin{itemize}
\item \textbf{A high-performance BLAS-based M2L implementation for \kifmm}, with algorithmic and layout optimizations that improve operational intensity, reduce precomputation time, and allow for simple implementations across architectures.
\item \textbf{A detailed benchmark comparison of FFT-based and BLAS-based M2L approaches}, using a clean, modular Rust implementation of the kernel-independent FMM, with performance evaluated across precision settings, setup \& runtime trade-offs, and architectural features. We restrict our benchmarks in this paper to single-node CPU architectures, to facilitate a comparison with \fftmtl.
\end{itemize}

We begin by briefly reviewing the \fmm and \kifmm in Section \ref{sec:fmm}, and describe the \mtl operation for the \kifmm in detail. In Section \ref{sec:blas_m2l} we describe the implementation of \blasmtl, discussing data layout and techniques to speed up the setup and runtime computations. In Section \ref{sec:benchmarks} we provide single node benchmarks. In Section \ref{sec:discussion} we compare the \blasmtl and \fftmtl approaches and discuss their relative merits, and conclude with a reflection on our results in Section \ref{sec:conclusion}.

\section{Fast Multipole Method}\label{sec:fmm}

\fmms rely on degenerate approximations of the kernel $K(\cdot, \cdot)$, such that the potential \eqref{eq:sec:introduction:potential} when evaluated between distant clusters of target and source points can be expressed as a sum,

\begin{equation}
    \phi_i \approx \sum_{p=1}^{P} \sum_{j=1}^N A_p(\xbfi)B_p(\ybfj)q_j, \> \> \> i=1,2...,M
    \label{eq:sec:fmm_review:degenerate_kernel}
\end{equation}

where the expansion order $P$ is chosen such that $P \ll, M, N$. The functions $A_p$ and $B_p$ are determined by the approximation scheme used by an \fmm variant. In the original presentation the evaluation of,

\begin{equation*}
    \hat{q}_{j, p} = \sum_{j=1}^N B_p(\ybfj)q_j, \> \> \> p=1,2,...,P
\end{equation*}

corresponded to the construction of an analytical expansion of the kernel function which represented the potential due to the set of source points \cite{Greengard1987}. The calculation,

\begin{equation*}
    \phi_{i} \approx \sum_{p=1}^P A_p(\xbfi) \hat{q}_{p, j}, \> \> \> i=1,2,...M
\end{equation*}

represented the evaluation of this potential at the set of target points.

The accuracy of the approximation \eqref{eq:sec:fmm_review:degenerate_kernel} depends on a sufficient distance between clusters of sources and targets, referred to as \textit{admissibility}. Therefore, sources considered in the second term of \eqref{eq:sec:introduction:near_far_split} are taken to correspond to admissible clusters, which can be approximated by \eqref{eq:sec:fmm_review:degenerate_kernel}, the near component evaluated directly.

Linear asymptotic complexity is achieved by partitioning the problem domain with a hierarchical data structure, an octree in three dimensions or a quadtree in two dimensions. A bounding box is placed over all sources and targets and recursively subdivided into equal sub-boxes, called its children. For a box $\sigma$ with side length $d$ centred at $c$, we define its near field $\mathcal{N}_\sigma$ as all boxes that lie within a box of side length $3d$ centered at $c$, including $\sigma$ itself. Its far field $\mathcal{F}_\sigma$ is the complement of this. The idea is then to compress evaluation of interactions for each target box where a source box can be considered admissible, ie. in its far field, by using approximations to represent the field generated by far-field boxes. The approximations for potential due to a set of source densities are encoded using \textit{multipole} and \textit{local} expansions. Multipole expansions are used to describe potentials in the exterior of given box generated from sources in that box, and local expansions are used to describe potentials generated by a source in a box which is considered in the far field of a given box. The terms `multipole' and `local' are common terminology across methods derived from the \fmm, even those which use an alternative approximation scheme. In octrees, adjacent boxes are those which share a face, edge, or vertex and are called \textit{neighbours}, the eight child boxes of a given box, are called \textit{siblings}. Translations between these expansions what gives the \fmm its complexity, in particular,
s

\begin{itemize}
    \item \textbf{Multipole to Multipole (M2M)}: Translation of the multipole expansion of a child box to one centered on its parent box. This allows boxes at coarser tree levels to accumulate a representation of the field of their descendent boxes as the tree is traversed from the finest to coarsest boxes level-by-level.
    \item \textbf{Multipole to Local (M2L)}: Translation of the multipole expansion of a source box, $\sigma$, into a local expansion of a non adjacent target box, $\tau$, whose parent is a neighbour of the source box's parent. Boxes $\sigma \in I_{\tau}$ are admissible for $\tau$ and are described as being in its \textit{interaction list}, $I_\tau$. In three dimensions $|I_\tau| \leq 189$. This operation summarizes the far field of a given box in a single local expansion.
    \item \textbf{Local to Local (L2L)}: Translation of a local expansion of a parent box, to one centered on each of its child boxes. This operation allows the transfer of the local expansion of a given box to its children as the tree is traversed from the coarsest to finest boxes level-by-level.
\end{itemize}

The algorithm for non-oscillatory kernels, based on \textit{uniform refinement} such that all leaf boxes are of the same size, proceeds in a recursive two step procedure.

\begin{enumerate}
    \item \textbf{Upward Pass}: First multipole expansions are encoded for boxes at the leaf level, in a \ptmfull step. We then recurse by level, from finest to coarsest boxes, applying the \mtm translation to each one.
    \item \textbf{Downward Pass}: The tree is then traversed from coarsest to finest boxes, the local expansion for each box is accumulated from both (i) its parent (\ltl) and (ii) source boxes in its interaction list (\mtl). At the leaf level, the local expansion is evaluated at target points in each leaf box (\ltp) and the potential due to adjacent boxes is calculated directly using (\ref{eq:sec:introduction:potential}) known as the \ptpfull operation.
\end{enumerate}

During the upward and downward passes each target box interacts with a fixed number of source boxes. Indeed, an octree for a set of points $N$ discretized such that each of leaf box contains a bounded number of points, results in $\sim N$ leaf boxes and depth $\log_8(N)$ contains $O(N)$ boxes in total, giving a runtime complexity of $O(\kappa N)$, where $\kappa$ is a constant that depends on the number of interactions for the \mtl and \ptp operations.

\subsection{Performance Characteristics of the FMM}\label{sec:fmm_review:sub:performance}

\fmm runtime is dominated by the \mtl and \ptp operations, and the trade-off between these operations is dictated by the depth of the octree. The independence of the sum \eqref{eq:sec:introduction:near_far_split} over target points means that the \ptp operation is easily expressed using either \acrshort{simd} or \acrshort{simt} paradigms. Furthermore, the reuse of source cluster data in \eqref{eq:sec:introduction:near_far_split} across multiple interactions with a target cluster leads to high arithmetic intensity, making this operation compute bound. On modern architectures using shallower trees results in a large near field calculations for each leaf box, and \ptp bound runtimes. We find that these can be performed quickly for moderate problems sizes on modern hardware as we demonstrate in Table \ref{tab:sec:introduction:direct} for \cpus tested in this work. The \mtl naturally results in non-contiguous memory access patterns and cache misses from having to index data corresponding to potentially discontinuous boxes in the tree. The fact that each box needs to perform up to 189 \mtl translations in three dimensions leads the \mtl operation to dominate runtime for deep trees.

We find that the performance of our \ptp implementation alleviates the need for \textit{adaptive refinement} of the hierarchical data structure, in which leaf boxes are refined until they contain fewer than a user-specified threshold of points. Previously adaptive \fmms have been recommended for highly non-uniform point distributions, and are used to limit the \ptp operation by increasing the number of admissible source boxes which may ostensibly be in the near field of a target box as defined above but are refined to a greater extent and therefore amenable to compression. However, this approach requires additional interaction lists which must be calculated at runtime from the point data \cite{Ying2004} as well as deeper trees to cope with the extreme point distributions. Ensuring contiguous data access is the most significant challenge when handling interaction lists and deeper trees make \fmms bound by the performance of the \mtl implementation. In our implementation we instead use a weaker form of adaptivity, whereby branches which contain no point data are pruned.

\begin{table}
    \centering
    \footnotesize
    \begin{tabular}{l l l l}
        \toprule
	    \textbf{Precision} & \textbf{Number of Points} & \textbf{Apple M1 Pro} & \textbf{AMD 3790X} \\
        \midrule
        \multicolumn{4}{c}{Single Threaded} \\
        \midrule
	    Single    & 5,000             & $11$ (ms)           & $11$ (ms) \\
		          & 20,000            & $180$ (ms)            & $171$ (ms)\\
        \midrule
	    Double    & 5,000             & $28.2$ (ms)           & $25$ (ms) \\
		          & 20,000            & $45.00$ (ms)            & $401$ (ms) \\
        \midrule
        \multicolumn{4}{c}{Multi Threaded} \\
        \midrule
	    Single    & 20,000             & $30$ (ms)             & $6 $ (ms) \\
		          & 100,000            & $820$ (ms)            & $148$ (ms)\\
		          & 500,000            & $21.80$ (s)             & $4$ ($s$) \\
        \midrule
	    Double    & 20,000             & $79$ (ms)           & $13$ (ms) \\
		          & 100,000            & $2.2$ (s)             & $308$ (ms) \\
		          & 500,000            & $56.1$ (s)            & $11$ (s) \\
        \bottomrule
    \end{tabular}
    \caption{Runtimes for single and multi threaded direct $O\left(N^2\right)$ evaluation of the Laplace kernel in the optimized implementation provided by our Green Kernels library \cite{bempp_green_kernels} for the \cpus listed in Table \ref{tab:sec:appendix:hardware_and_software}. We consider the source and target points to be the same set. Our implementation of the Laplace kernel closely follows that first presented in \cite{Malhotra2015}, using explicit SIMD programming and Newton iterations for the calculation of fast square roots in the kernel function.}
    \label{tab:sec:introduction:direct}
\end{table}

\subsubsection{Forming Multipole and Local Expansions}\label{sec:introduction:sub:kifmm_review:sub:forming_expansions}

\begin{figure}[h]
    \centering
    \includegraphics[width=0.7\textwidth]{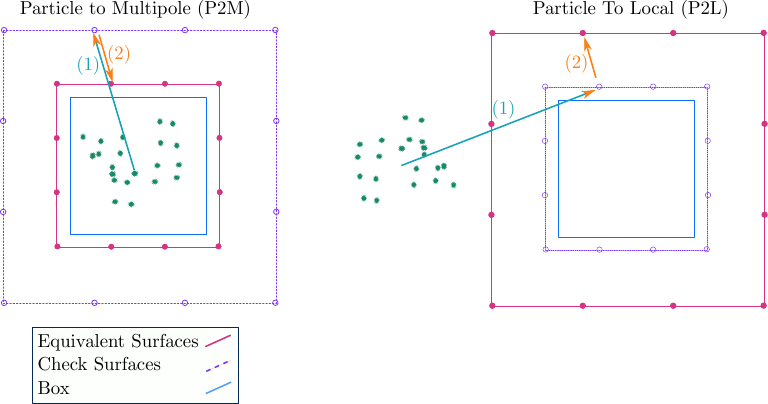}
    \caption{We illustrate the surfaces, and their associated discretization points, required to construct multipole and local expansions from source points, shown in green. This is for a problem in $\rthree$, where we have taken $P=4$. We show cross sections of cubic surfaces, as in Figure 4 of \cite{Ying2004}, where source points are shown as green points. The arrows illustrate the steps of the calculation: (1) compute the check potential from the source points,  (2) calculate the equivalent densities using the check potentials.}
    \label{fig:p2m}
\end{figure}

We review the \kifmm of Ying. et. al \cite{Ying2004}, which makes use of the \mfsfull for field approximations. Consider the construction of a `multipole' expansion corresponding to a set of source densities in a box $\sigma$, as in the left plot of Figure \ref{fig:p2m}. We begin by constructing an `equivalent surface', $\ybf^{\sigma, u}$, represented by discrete points $\{\ybf_j\}_{j=1}^{N_\text{equiv}}$, and are associated with weights called `equivalent densities', $q^{\sigma, u}$, with $u$ signifying that we are talking about `upwards surfaces' for the construction of multipole expansions. We match the field generated by these to that generated by the true source densities at a further enclosing `check surface', $\xbf^{\sigma, u}$ represented by discrete points $\{\xbf_i \}_{i=1}^{N_{\text{check}}} \in \xbf^{\sigma, u}$ to arrive at the equation,

\begin{equation}
   \sum_{j \in I_s^{\sigma}} K(\xbf_i, \ybfj) q_j = \phi^{\sigma, u}(\xbf_i),~\xbf_i\in \xbf^{\sigma,u}
\end{equation}

where $\phi^{\sigma, u}$ is the (upward) `check potential', and $I_s^{\sigma}$ is the index set of the sources contained in $\sigma$, $\{\ybfj\}_{j \in I_s^{\sigma} }$ and their associated densities, $\{q_j\}_{j \in I_s^{\sigma}}$. This gives the matrix equation,

\begin{equation}
    \label{eq:check_potential_matvec}
        \mathbsf{K} \mathbsf{q} = \bm{\Phi}
\end{equation}

where $\mathbsf{K}$ is an $N_{\text{check}} \times N_{\text{equiv}}$ matrix with entries $[\mathbsf{K}]_{i, j} = K(\xbfi, \ybfj)$ with $\{\xbf_i\}_{i=1}^{N_{\text{check}}}$ the evaluation points on the check surface and $\{\ybf_j\}_{j=1}^{N_{\text{equiv}}}$ the source points on the equivalent surface, where $N_{\text{check}}$ and $N_{\text{equiv}}$ are the number of evaluation points on the check surface and source points on the equivalent surface respectively. We associate the equivalent densities $\mathbsf{q}$ with a multipole expansion describing the potential due to sources contained in $\sigma$. Equation \eqref{eq:check_potential_matvec} defines an ill-conditioned least-squares problem, which we solve using the `backward-stable' pseudo inverse first presented in \cite{Malhotra2015} to find the equivalent densities $\mathbsf{q}$ \footnote{In \cite{Malhotra2015} they note there are two sources of error in computing equivalent densities. The first is a round-off error in the check potential, which is amplified by a factor equal to the condition number of the inverted matrix. The second is from multiplying the factors of the pseudo inverse itself together, due to the large range in the singular values. They state that the first error is damped by a similar factor when computing a far field potential using an equivalent density, and therefore does not significantly effect the final computed potential using the \kifmm. However, the second error source can be mitigated by storing the pseudo inverse in two components where the diagonal matrix of singular values is multiplied with one of the two orthonormal matrices of left or right singular vectors}.

\begin{flalign}
    \label{eq:sec:solve_pinv}
    \mathbsf{q} &= (\mathbsf{K})^\dag \bm{\Phi} \\
    &= \mathbsf{V}_{\varepsilon} \bm(\Upsigma)_{\varepsilon}^{-1} \mathbsf{U}^T_{\varepsilon} \bm{\Phi} \\
    &= \mathbsf{V}_{\varepsilon} \tilde{\mathbsf{U}}^T_{\varepsilon} \bm{\Phi}
\end{flalign}

Greater accuracy is achieved by storing separately the components of the pseudo inverse, where $\varepsilon$ is a threshold for singular values beyond which associated singular vectors are filtered out. Our pseudo inverse implementation closely mirrors that of SciPy \cite{virtanen2020scipy}. Considering a matrix of size $M \times N$ for which we are computing our pseudo inverse defining $d = \max(M, N)$, $\alpha$ (absolute tolerance) as a user specified value, $\rho$ (relative tolerance) as $\rho = d \cdot \varepsilon_{\text{mach}}$ where $\varepsilon_{\text{mach}}$ is machine precision and letting $\sigma_0$ represent the largest singular value, the threshold $\varepsilon$ is computed as,

\begin{equation}
    \varepsilon = (\alpha + \rho) \cdot \sigma_0.
\end{equation}

We find good performance hueristically by setting $\alpha = 0$.

One can construct local expansions using the surfaces defined in the right plot of Figure \ref{fig:p2m}, where now the check surface must be enclosed by the equivalent surface.

\subsection{Translation Operators}

\begin{figure}[h]
    \centering
    \includegraphics[width=\textwidth]{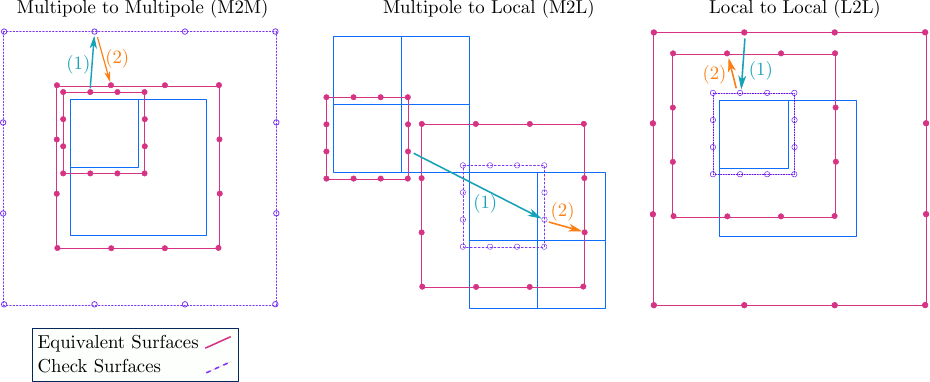}
    \caption{We illustrate the surfaces, and their associated discretization points, required to perform each field translation, \mtm, \mtl and \ltl. This is for a problem in $\rthree$, where we have taken $P=4$. We show cross sections of cubic surfaces, as in Figure 5 of \cite{Ying2004}. We show how for the \mtl, the source box's equivalent surface and the target box's check surface can be seen to originate from a regular Cartesian grid, enabling the acceleration in the computation of the check potential in this case via the \fft. The arrows illustrate the flow of the calculation: (1) compute the check potential, (2) evaluate the equivalent density using the check potential.}
    \label{fig:field_translations}
\end{figure}

The \mtm, \ltl and \mtl operators are formed using the same technique as above, and are illustrated in Figure \ref{fig:field_translations}. For the \mtm and \ltl, forming the required surfaces, we match the field generated by the child/parent box's equivalent densities at the parent/child box's check surface respectively to find the check potential. This results in equations of the form of \eqref{eq:sec:solve_pinv} for the calculation of the the multipole/local expansions, respectively.

We explicitly show the procedure for the \mtl operator. Denote by $\xbfi$ the points discretizing the downward check surface of the target box $\tau$, and by $\ybfj$ the points discretizing the downward equivalent surface of a source box in its interaction list, $\sigma \in I_\tau$. We first need to evaluate the field generated at $\xbfi$ through the equivalent surface the source box. This can be written as a matrix product,

$$
\bm{\Phi}^{\sigma, \tau} = \mathbsf{K}_{\sigma, \tau} \mathbsf{q}^{\sigma}
$$

with $[\mathbsf{K}_{\sigma, \tau}]_{i, j} = K(\xbfi, \ybfj)$ and the vector $\mathbsf{q}^\sigma$ containing the associated equivalent source densities in $\sigma$. Introducing the matrix $\mathbsf{K}_\tau$, which contains the kernel interactions between the discretization points on the equivalent surface and the check surface for $\tau$, the equivalent density $\mathbsf{q}^\tau$ is obtained through,

\begin{equation}
    \label{eq:m2l_matvec}
    \mathbsf{q}^\tau = (\mathbsf{K}_\tau)^\dagger \sum_{\sigma \in I_\tau} \mathbsf{K}_{\sigma, \tau} \mathbsf{q}^\sigma,
\end{equation}

where the sum is over all boxes in $\sigma$ in the interaction list $I_\tau$ of $\tau$.

The calculation \eqref{eq:m2l_matvec} corresponds to a series of dense matrix-vector products, where the matrix $\mathbsf{K}_{\sigma, \tau}$ is known to be of low rank due to the asymptotic smoothness of the kernel. They are therefore amenable to numerical compression techniques such as the \svd. Alternatively, for kernels which are also translation invariant such that $K(\xbf, \ybf) = K(\xbf-\mathbsf{c}, \ybf - \mathbsf{c})$ for $\mathbsf{c} \in \mathbb{R}^d$, by choosing the upward equivalent and downward check surfaces to be defined equivalently with respect to a given box, the evaluation of the check potential can be interpreted as a three dimensional convolution and can therefore be accelerated with a \fftfull. We note that $(\mathbsf{K}_\tau)^\dagger$ depends only on the downward equivalent and check surfaces of a box, if these are chosen to be the same relative to a given box, this matrix can be cached, and depending on kernel properties, scaled at each level of the tree.

\section{Blas Based M2L Operators For the Laplace Kernel}\label{sec:blas_m2l}

\subsection{Algorithm Setup}\label{sec:blas_m2l:sub:algorithm_setup}

To accelerate the evaluation of \eqref{eq:m2l_matvec} we compress the matrices $\mathbsf{K}_{\sigma, \tau}$  associated with each \mtl translation numerically, which we call the \mtl-matrices. We build upon ideas first presented in \cite{Fong2009,messner2012optimized} for the \bbfmm. Each source/target box pair corresponds to a relative orientation between the boxes which we describe with a transfer vector $t \in T_l$ where $T_l$ is the set of all unique transfer vector at a level $l \in [0, d]$ of an octree of depth $d$.

We note that the size of the set $|T_l| \leq 316$ at a given level $l$ of an octree for the Laplace kernel. This is seen by considering the interactions of a given target box, $\tau$, and its siblings together. Each sibling box's near field consists of $27^3$ boxes including itself, with which it shares an edge, a face or a vertex. From the definition of admissible boxes, a given target box's far field consists of $189=6^3 -3^3$ boxes. Where $6^3$ comes from the fact that the interaction list of each box must lie within the near field of $\tau$'s parent, which has a width of 6 boxes in units of its child boxes. Similarly, the union of all possible relative positions corresponding to interactions between each sibling and child boxes in the near field of its parent can be seen to be $7^3$. Accounting for the relative positions for near field interactions which are common to each sibling, we see that there are at most $316=7^3-3^3$ unique relative positions for translationally invariant kernels for which $K(\xbf - \ybf) = K(\xbf, \ybf)$, such as the Laplace kernel.

We identify each \mtl-matrix with a transfer vector $\mathbsf{K}_t$, and assemble a $N_{\text{check}} \times |T_l| \cdot N_{\text{equiv}}$ matrix row-wise,

$$
\mathbsf{K}_{\text{fat}} = \begin{bmatrix} \mathbsf{K}_1 & ... & \mathbsf{K}_{|T_l|} \end{bmatrix}
$$

and a $|T_l| \cdot N_{\text{check}} \times N_{\text{equiv}}$ matrix column-wise

$$
\mathbsf{K}_{\text{thin}} = \begin{bmatrix} \mathbsf{K}_1\\ ... \\ \mathbsf{K}_{|T_l|} \end{bmatrix}.
$$

These are then compressed using \acrshortpl{svd}, providing a rank-$k$ approximation,

\begin{flalign}
    \label{eq:sec:optimizations:k_fat}
    \mathbsf{K}_{\text{fat}} &\approx \mathbsf{U} \bm{\Upsigma} \begin{bmatrix} \mathbsf{V}^{T}_1 ... \mathbsf{V}^{T}_{|T_l|} \end{bmatrix} = \mathbsf{U} \bm{\Upsigma} \tilde{\mathbsf{V}}^T, \\
    \label{eq:sec:optimizations:k_thin}
    \mathbsf{K}_{\text{thin}} &\approx \begin{bmatrix} \mathbsf{R}_1\\ ...\\ \mathbsf{R}_{|T_l|} \end{bmatrix}  \bm{\Uplambda} \mathbsf{S}^T = \tilde{\mathbsf{R}} \bm{\Uplambda} \mathbsf{S}^T,
\end{flalign}

$\mathbsf{U}$ is of size $N_{\text{check}} \times k$ and $\mathbsf{S}$ is of size $N_{\text{equiv}} \times k$ and $k$ is chosen such that we achieve a desired error $\varepsilon$ in the evaluated potential.

From \cite{Fong2009}, considering each element of $\mathbsf{K}_{\text{fat}}$ and $\mathbsf{K}_{\text{thin}}$ corresponding to a given $t \in T_l$, and using that $\mathbsf{S}$ and $\mathbsf{U}$ have orthonormal columns, we obtain

\begin{equation}
    \mathbsf{K}_t = \mathbsf{U} \mathbsf{C}_t \mathbsf{S}^T,
\end{equation}

where $$\mathbsf{C}_t = \mathbsf{U}^T \mathbsf{K}_t \mathbsf{S}$$ is of size $k \times k$ and is called the \textit{compressed \mtl-matrix}, which operates on the \textit{compressed multipole expansion}, $$\tilde{\mathbsf{q}} = \mathbsf{S}^T \mathbsf{q} $$ and returns the \textit{compressed check potential} $$\tilde{\bm{\phi}} = \mathbsf{C}_t \mathbsf{S}^T \mathbsf{q} = \mathbsf{C}_t \tilde{\mathbsf{q}} $$

Examining the singular values of $\mathbsf{K}_{\text{fat}}$ and $\mathbsf{K}_{\text{thin}}$ of dimension $M \times N$ for the three dimensional Laplace kernel one observes that approximately $k \sim \max{(M , N)}/2$ is sufficient for obtaining machine precision relative errors in both single and double precision in the final evaluated potential as shown in \cite{Fong2009}. We find in practice that we can obtain significantly greater compression than this when solving with for a lower level of accuracy as we discuss in Section \ref{sec:benchmarks}. For kernels which are symmetric such that $\mathbsf{K}_{\text{fat}} = \mathbsf{K}_{\text{thin}}^T$ where the check and equivalent surfaces are discretized to the same degree we can compute just a single \svd. If a kernel is \textit{homogenous} such that $K(\alpha \xbf, \alpha \ybf) = \alpha K(\xbf, \ybf)$ such as the Laplace kernel we see that we need only compute the \svd \textit{once}, with the resulting \mtl matrices reused and scaled at each level.

Messner et al notice that as the value of $k$ is dictated by the highest rank interaction in $T_l$ the cost of the applying $\mathbsf{C}_t$ can be further reduced with another \svd for each individual $\mathbsf{K}_t$ for $t \in T_l$ \cite{messner2012optimized},

\begin{equation}
    \label{eq:directional_compression}
    \mathbsf{C}_t \approx \mathbsf{\bar{U}}_t \bar{\bm{\Upsigma}}_t \mathbsf{\bar{V}}_t^T = \mathbsf{\bar{U}}_t \mathbsf{\bar{V}}_t'^T
\end{equation}

where $ \mathbsf{\bar{U}}_t $ and $\mathbsf{\bar{V}}_t'$ are of size $k \times k_t$ and $\bar{\bm{\Upsigma}}$ is of size $k_t \times k_t$ and $k_t$ again chosen to preserve a given error $\varepsilon$ in the evaluated potential calculated via \eqref{eq:l2_error}. In order to determine $k$ and $k_t$ we perform a grid search to determine the optimum parameters to achieve a given error, the results of this are described in Table \ref{tab:app:grid_search_laplace_blas_m2l} in Appendix \ref{app:optimal_parameters}.

In our \blasmtl approach we simply identify \textit{all} matching source/target box pairs in each level of an octree $l \in [2, d]$ during the downward pass by their transfer vector, grouping together the computation of their check potentials as a \textit{single matrix-matrix product} for each transfer vector at each level $l$. The large size of these matrix-matrix products is naturally of high operational intensity.

This is in contrast to previous \blasmtl schemes such as that described in \cite{messner2012optimized}, where the symmetry and translational invariance properties of the kernel are used to permute the \mtl matrices for a given source/target box pair to form matrix-matrix products. Further blocking was accomplished by blocking over sets of sibling boxes.

With this we can compute all check potentials for target boxes in a given level in at most $2 \times |T_l| = 634$ level 3 \blas calls for translationally invariant kernels. High performance is enabled by the powerful optimizations implemented by modern \blas variants \cite{VanZee2015,OpenBLAS,NVIDIA_cuBLAS}, automatically configuring tiling schemes for the memory hierarchy of a particular device and applying optimizations such as loop re-ordering and \simd vectorisation.

Computing the \acrshortpl{svd} required by our scheme is is costly in comparison to the setup of the \fftmtl, especially for high expansion orders in double precision. We address this by using the \rsvdfull \cite{halko2011finding}, in contrast the deterministic \acrshortpl{svd} used in the past \cite{messner2012optimized}. We only consider `one shot' \acrshortpl{rsvd} that avoid slow power iterations which repeatedly apply \acrshort{qr} decompositions to compute the orthonormal basis of the subspace into which the matrix to be compressed is being projected, instead only relying on oversampling and our knowledge of the low-rank nature of our kernel to improve the performance of the \rsvd. Specifically, we use the result that $\mathbsf{K}_{\text{fat}}$ and $\mathbsf{K}_{\text{thin}}$ is of dimension $M \times N$ are approximately of rank $k \sim \max{(M, N)}/2$ for the Laplace kernel as found in \cite{Fong2009}, computing the \rsvd with this rank estimate. This estimate must be updated accordingly for each kernel considered.

We note that the \acrshortpl{svd} required during the directional compression step \eqref{eq:directional_compression} are relatively cheap in comparison to that for $\mathbsf{K}_{\text{fat}}$ and $\mathbsf{K}_{\text{thin}}$ and therefore use only the classical \svd here while retaining reasonable setup times, which we document in Table \ref{tab:app:grid_search_laplace_blas_m2l}.

In our current implementation we perform the \rsvd using the linear algebra and \acrshort{blas} libraries listed in Table \ref{tab:sec:appendix:hardware_and_software} for each architecture, and compute the classical \acrshortpl{svd} required by the directional compression steps in a parallel loop over each direction.

\subsection{Algorithm}\label{sec:sub:algorithm}

We summarize our \blasmtl scheme at each level $l \in [2, d]$ during the downward pass of the \kifmm,

\begin{enumerate}
    \setlength\itemsep{1em}
    \item For each transfer vector (associated with index $t$) collect all corresponding source/target box pairs as a map. This step is a part of the precomputation. Let $N_t$ be the number of source boxes associated with $t$. By $N_\sigma$ we denote the number of all source boxes in the tree at level $l$.
    \item For all $N_\sigma$ source boxes compute the compressed multipole expansions. This is done through a single \acrshort{blas}3 call, \[ [\tilde{\mathbsf{q}}_1,...,\tilde{\mathbsf{q}}_{N_{\sigma}}] = \mathbsf{S}^T [\mathbsf{q}_1,..., \mathbsf{q}_{N_{\sigma}}]\]
    \item Allocate up to 316 buffers to store each of the $N_t$ compressed multipole expansions associated with each transfer vector $t \in [1, 316]$ contiguously, noting that not all transfer vectors may appear in interaction lists at a given level. Each buffer is of the form,
        \[ [\tilde{\mathbsf{q}}_1,...,\tilde{\mathbsf{q}}_{N_t}] \]
    This requires using the results of Step 1, where we identify transfer vectors associated with each source box, which can be used to lookup their associated data.
    \item Compute the compressed check potentials in a loop over each $t \in [1, 316]$, resulting in up to $632$ \acrshort{blas}3 calls, as not all transfer vectors may appear in interaction lists at a given level,
        \[ [\bm{\xi}_1, ..., \bm{\xi}_{N_t}] = \bar{\mathbsf{V}}_t'^T [\tilde{\mathbsf{q}}_1,...,\tilde{\mathbsf{q}}_{N_t}]  \]
        \[ [\tilde{\bm{\phi}}_1, ..., \tilde{\bm{\phi}}_{N_t}] = \bar{\mathbsf{U}}_t  [\bm{\xi}_1,...,\bm{\xi}_{N_t}]  \]

    Where we have used the re-compressed form of $\mathbsf{C}_t$ calculated with \eqref{eq:directional_compression}. We note that in implementations these two multiplications can be performed in place, and that the product $\bar{\mathbsf{V}}_t'^T = \bm{\bar{\Sigma}}_t \mathbsf{\bar{V}}_t^T$ should be precalculated.

    \item Each $\tilde{\bm{\phi}}_i$ is associated with a source box at level $l$ i.e. contained in an interaction list of any target box at this level. These are accumulated in a buffer containing $N_{\tau}$ compressed check potentials associated with each target box at level $l$. We now compute with another \acrshort{blas}3 call the corresponding uncompressed check potentials, \[ [\bm{\phi}_1,...,\bm{\phi}_{N_{\tau}}] = \mathbsf{U} [\tilde{\bm{\phi}}_1, ..., \tilde{\bm{\phi}}_{N_{\tau}}] \]
\end{enumerate}

Once the check potentials are found, we can recover the local expansions at each target box using a calculation of the form of \eqref{eq:m2l_matvec}.

The entire scheme for computing the check potentials at each tree level $l$ during the downward pass, consists of up to $634$ matrix-matrix multiplications per level, including the conversion into the compressed multipole expansion, the calculation of the compressed check potentials, and the conversion of the compressed check potential back into the check potential. If we choose to avoid the re-compression of $\mathbsf{C}_t$ during Step 4 the scheme uses 318 matrix-matrix multiplications per level.

The organisation of matrix-matrix multiplications in Step 4 is up to the choice of an implementer. In a \cpu implementation, if a particular architecture contains a large number of \cpu cores we find good performance by partitioning the multiplications for each $t \in [1, 316]$ over threads and performing them each in single threaded mode. In this case instruction level and data prefetching optimizations of the \blas library are still enabled. Conversely if a \cpu architecture has limited numbers of cores such as on standard desktops and laptops we find good performance with the multiplications themselves permed multithreaded while we loop serially over each transfer vector.

The only runtime data organisation required is the allocation of buffers in Step 3, and the accumulation of data for each target box in Step 5. Both of these steps can be parallelised with multithreading, as data for each source and target box are not overlapping in memory. The scheme is also readily adapted into a \gpu based implementation, especially if a particular architecture supports coherent memory access between the \cpu and \gpu. The data organisation can be made to take place on the \cpu with the matrix-matrix multiplications deferred to the \gpu with batched-\acrshort{blas} \cite{NVIDIA_cuBLAS}, with minimal data transfer costs.

We defer to the underlying \acrshort{blas} implementation to apply instruction level optimizations, and maximize arithmetic intensity. Instead, we simply ensure that the data accesses required by the \acrshort{blas} calls are contiguous. Importantly, this approach allows us to easily compute the FMM for \textit{multiple sets of source densities} sharing a set of target and source points, common in the application of \fmms to boundary integral equations. We simply identify all common translations corresponding to each right hand side of \eqref{eq:sec:introduction:potential} at a given level, and pass them through the level 3 \acrshort{blas} operations, letting the underlying \acrshort{blas} library handle the required blocking for this larger calculation. We provide an algorithmic analysis of the runtime in Appendix \ref{app:analysis} to find estimates for the operational intensity of our method. The key point to note is that the operational intensity of the convolution step of our \blasmtl method in Step 4 in which we calculate the compressed check potentials in the above algorithm from \eqref{eq:app:operational_intensity_convolution_laplace} is given as,

\begin{equation*}
   \frac{2 \cdot 8^l \cdot k \cdot k_t}{ 8^l \cdot k + k_t \cdot k + 2 \cdot  8^l \cdot k_t } + \frac{2 \cdot 8^l \cdot k \cdot k_t }{8^l \cdot k_t + k_t \cdot k + 2 \cdot 8^l \cdot k}  \> \> \text{FLOPs/Accesses}.
\end{equation*}

Assume for simplicity that $k = k_t$. Then the above expression simplifies to,

\begin{equation*}
    \frac{4 \cdot 8^l k}{3 \cdot 8^l + k},
\end{equation*}

which shows that FLOPs/Accesses increases with growing $k$, which is especially advantageous for the higher ranks in three dimensional \fmms.

\section{Benchmarks}\label{sec:benchmarks}

We test our software on the two \cpu architectures listed in Table \ref{tab:sec:appendix:hardware_and_software}. The x86 AMD 3790X architecture provides an example of a commodity high performance \cpu with a high core count and large cache memory and \acrshort{ram}. This architecture allows us to test the performance of our shared memory optimizations across large numbers of threads on a single node. The ARM Apple M1 Pro architecture though not typically used in \acrshort{hpc} benchmarks gives an interesting comparison both in terms of the underlying ARM-\acrshort{isa} but additionally of an architecture supporting a large and highly efficient `unified memory' pool with atypically large cache sizes. Furthermore, it has a core count typical of machines available to the majority of developers. The M1 Pro's unified memory supports significantly higher computational throughput than the AMD architecture, and allows us to observe the impact of operational intensity on runtimes with respect to the more typical AMD architecture. Interestingly, when using Apple's Accelerate \cite{AppleAccelerate} framework for \acrshort{blas} operations the M1 Pro makes use of specialized registers for matrix computations, offering an example of an architectural feature which can be automatically exploited by the structure of our algorithm in terms of \acrshort{blas} operations when using the \blasmtl approach. We also note that such specialised matrix registers aren't unique to Apple's M Series and are also becoming more common in architectures from other vendors.

Our benchmark problems are the evaluation of potentials between $1 \times 10^6$ and $8 \times 10^6$ source and target points distributed uniformly in a unit box. Sources and targets are considered to be the same set, with sources assigned random source densities. We approximate error in the evaluated potential, $\varepsilon$, with the $L_2$ norm of the the error in the \fmm approximation,

\begin{equation}
    \label{eq:l2_error}
    \varepsilon = \left ( \frac{\sum_{i=1}^M |\phi_i - \tilde{\phi}_i|^2 }{\sum_{i=1}^M |\phi_i|^2 } \right )^{1/2},
\end{equation}

where $\tilde{\phi}_i$ are the potentials obtained via the \fmm and $\phi_i$ are those computed by direct calculation with the same precision at each target point $\{\xbfi\}_{i = 1}^M$. We identify via grid-search the parameter settings for the \fmm which allow us to achieve a given error of $\bigO{\varepsilon}$ in the least time. The results of this are shown in Tables \ref{tab:app:grid_search_laplace_fft_m2l} and \ref{tab:app:grid_search_laplace_blas_m2l} when using the \fftmtl and \blasmtl approaches respectively. Here, we also break down the setup costs when using either \blasmtl and \fftmtl by the time required to construct octrees, allocate required buffers and set up index pointers for tree traversal, precomputing maps to lookup data at runtime as well as the precomputation cost of both \mtl approaches. We observe a significant variation in the optimal parameters between both architectures. On the AMD architecture we are typically able to take shallower trees, where leaf boxes contain in the order of 1000s of particles, likely due to its higher core-count for which our \simd based \ptp implementation is highly optimized.

On the AMD architecture we use OpenBLAS for linear algebra and Apple Accelerate on the M1 Pro. We note that Apple Accelerate is able to significantly reduce setup times time when using \blasmtl in contrast to OpenBLAS, especially at high expansion orders in double precision despite the fewer computational resources of the M1 Pro in comparison to the AMD 3790X. We observe that for low accuracy experiments, especially in single-precision the overall setup times when using either \fftmtl or \blasmtl are broadly comparable, however the cost of the matrix compressions required by \blasmtl begins to grow significantly at the highest accuracies tested in double precision.

The runtimes when using \blasmtl and \fftmtl are illustrated in Table \ref{tab:benchmarks-m1} for the M1 Pro and Table \ref{tab:benchmarks-amd} for the AMD architecture respectively. We illustrate the corresponding computational throughputs and required storage of the \mtl operation when using either \blasmtl or \fftmtl in Tables \ref{tab:throughput-m1} and Table \ref{tab:throughput-amd}, respectively.

In single precision we don't observe a significant difference in runtimes when using \blasmtl and \fftmtl on either of the architectures tested for either problem size. However, at the lowest accuracies tested the \fftmtl consistently outperforms the \blasmtl - likely due to its formulation with high arithmetic intensity (See Appendix \ref{sec:fft_m2l_analysis}), and the relatively small data sizes of the \mtl operator as seen in Tables \ref{tab:throughput-m1} and \ref{tab:throughput-amd}. However, we notice from Tables \ref{tab:throughput-m1} and \ref{tab:throughput-amd} that the \fftmtl approach rapidly achieves bandwidth saturation when the associated data sizes of the \mtl operator matrices is larger than, or comparable in size to, the L2 cache of each architecture. This results in diverging performance for high-accuracy experiments in double precision, with evaluations based on the \blasmtl outperforming those using \fftmtl on both architectures tested due to its increasing operational intensity for larger matrix sizes. Though we note that the effect of this diverging performance is less pronounced for larger problem sizes, where the \ptp operation dominates, especially on the AMD architecture with its less favorable memory architecture.

The unified memory architecture of the M1 Pro, as well as its specialized matrix processing registers, make it interesting to see how it handles multiple source density vectors for a given point distribution when using the \blasmtl approach. We observe in Table \ref{tab:benchmarks-m1} the benefit of further increased operational intensity on per-vector end-to-end \kifmm runtimes, especially in single precision, when using multiple source density vectors. This effect is visibly reduced on the AMD architecture shown in Table \ref{tab:benchmarks-amd}, where even in single precision the larger data sizes often lead to deteriorating performance by increasing the number of source density vectors.

\begin{figure}[!htbp]
    \centering 
    \includegraphics[width=.6\linewidth]{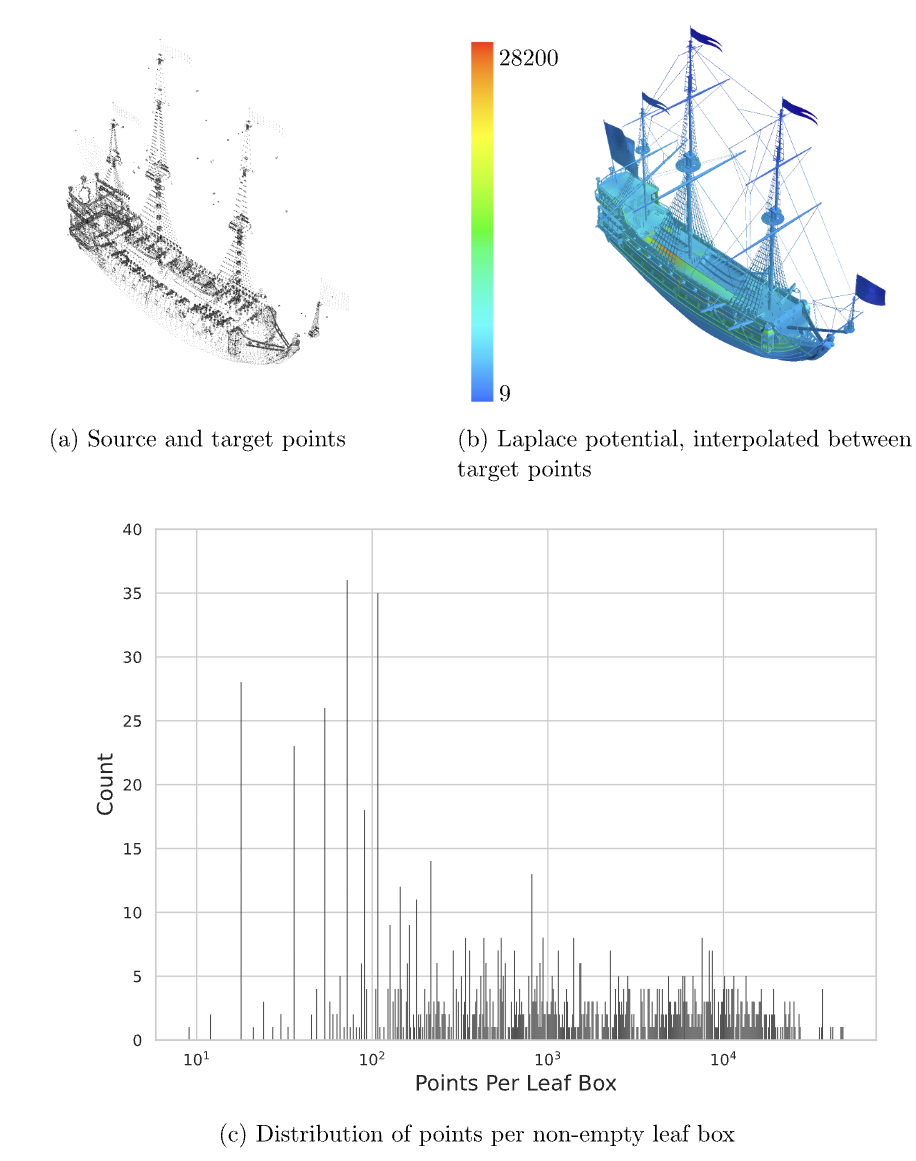} 
    \caption{
        The Laplace potential is computed in double precision such that the error, calculated using \eqref{eq:l2_error}, is $\bigO{\varepsilon} = 10^{-10}$. We compare \fftmtl and \blasmtl on a highly non-uniform point distribution consisting of 1,441,572 vertices from 480,524 triangles, each with unit source density. The octree is uniformly refined to a depth of $d=5$, producing 2,184 non-empty leaf boxes. The same set of points is used as both sources and targets.
        \fftmtl is run with equivalent surface order $P_e = 10$ and a batch size of 16. \blasmtl is configured with $P_e = 10$, check surface order $P_c = 11$, an singular value threshold of $1 \times 10^{-11}$ below which associated singular vectors in the compressed \mtl matrices are cutoff, and 20 oversamples in the \rsvd.
        The mean runtime on this non-uniform dataset is 5,779 ms for \fftmtl and 5,176 ms for \blasmtl. For comparison, the same number of uniformly distributed source and target points (with identical FMM parameters and octree depth $d = 4$ yielding a similar level of accuracy - the lower level of refinement used to better handle the uniform distribution) yield runtimes of 3,933 ms (\fftmtl) and 4,749 ms (\blasmtl). Thus, the non-uniformity incurs a $\sim$ 47\% runtime increase for \fftmtl and $\sim$ 8\% for \blasmtl compared to the evaluation over a uniform point distribution which can be interpreted as a baseline.
        The histogram illustrates the uneven point distribution, with leaf boxes containing between 1 and $10^5$ points. Geometry from \cite{thingiverse_3253610}.
    }
    \label{fig:sec:benchmarks:ship}
\end{figure}

In Figure \ref{fig:sec:benchmarks:ship}, we evaluate the effectiveness of our approach for computing potentials in scenarios involving a highly non-uniform point distribution. We observe a relatively modest increase in runtime in this case compared to a uniform random distribution with the same number of source and target points evaluated to a similar level of accuracy in the final potential taken as a baseline ($\sim$ 47 \% when using \fftmtl and $\sim$ 8 \% when using \blasmtl on the test problem illustrated in Figure \ref{fig:sec:benchmarks:ship}). This demonstrates the suitability of our uniform tree implementation for tackling non uniform point distributions arising from the efficiency of our \ptp implementation, and memory intensive operations confined to a minimal number of \mtl operations.

\begin{table}[htbp]
  \centering
  \caption{Mean end-to-end algorithmic ($\mathbsf{T}_{\text{FMM}}$), M2L ($\mathbsf{T}_{\text{M2L}}$), and P2P ($\mathbsf{T}_{\text{P2P}}$) runtimes in milliseconds on the Apple M1 Pro for achieving an $L_2$ relative error $\bigO{\varepsilon}$ calculated using \eqref{eq:l2_error} using both \blasmtl and \fftmtl when evaluating the potentials between $N$ source and target points, taken to be the same set. Runtimes are reported for the best parameter settings shown in Tables~\ref{tab:app:grid_search_laplace_fft_m2l} and \ref{tab:app:grid_search_laplace_blas_m2l}, with error reported from the final digit. For \blasmtl, values in brackets indicate the number of source density vectors, where we report the runtimes per-vector. The fastest mean end-to-end runtimes for each error threshold are highlighted. We note that for $N=8 \times 10^6$, the single precision results saturate at an $L_2$ relative error of $\bigO{10^{-3}}$ on the M1 Pro. This is likely due to a combination of aggressive single-precision optimizations applied by Apple Accelerate, accumulated roundoff error from the larger number of closely spaced points, and architecture-specific \acrshort{simd} reduction ordering.}
  \label{tab:benchmarks-m1}
  {\scriptsize
  \begin{tabular}{l l l l l l l l l l l l l}
    \toprule
    & \multicolumn{3}{c}{\textbf{\blasmtl (1)}} & \multicolumn{3}{c}{\textbf{\blasmtl (5)}} & \multicolumn{3}{c}{\textbf{\blasmtl (10)}} & \multicolumn{3}{c}{\textbf{FFT-M2L}} \\
    $\bigO{\bm{\varepsilon}}$ & $\mathbsf{T}_{\text{FMM}}$ & $\mathbsf{T}_{\text{M2L}}$ & $\mathbsf{T}_{\text{P2P}}$ & $\mathbsf{T}_{\text{FMM}}$ & $\mathbsf{T}_{\text{M2L}}$ & $\mathbsf{T}_{\text{P2P}}$ & $\mathbsf{T}_{\text{FMM}}$ & $\mathbsf{T}_{\text{M2L}}$ & $\mathbsf{T}_{\text{P2P}}$ & $\mathbsf{T}_{\text{FMM}}$ & $\mathbsf{T}_{\text{M2L}}$ & $\mathbsf{T}_{\text{P2P}}$ \\
    \midrule
    \multicolumn{13}{c}{$\bm{N = 1 \times 10^6$}} \\
    \midrule
    \multicolumn{13}{c}{\textbf{Single Precision}} \\
    \midrule
	  $10^{-3}$ & 196 & 107 & 80 & 162 & 76 & 82 & 153 & 67 & 83 & \boldcolor{darkgreen}{152} & 64 & 80 \\
	  $10^{-4}$ & 243 & 147 & 80 & 211 & 116 & 83 & \boldcolor{darkgreen}{203} & 109  & 85 & 256 & 163 & 80 \\
    \midrule
    \multicolumn{13}{c}{\textbf{Double Precision}} \\
    \midrule
	  $10^{-6}$ & 741 & 499 & 158 & \boldcolor{darkgreen}{686} & 460 & 159 & 705 & 483 & 163 & 1061 & 813 & 160 \\
	  $10^{-8}$ & 1293 & 875 & 159 & \boldcolor{darkgreen}{1286} & 880 & 161  & 1331 & 932 & 161 & 1477 & 257 & 1101 \\
	  $10^{-10}$ & 1563 & 237 & 1096 & \boldcolor{darkgreen}{1502} & 227 & 1077 & \boldcolor{darkgreen}{1502} & 222 & 1084 & 1799 & 482 & 1086 \\
    \midrule
    \multicolumn{13}{c}{\textbf{$\bm{N = 8 \times 10^6}$}} \\
    \midrule
    \multicolumn{13}{c}{\textbf{Single Precision}} \\
    \midrule
    $10^{-3}$ & 1635 & 891 & 668 & 1432 & 651 & 668 & 1308 & 596 & 662 & \boldcolor{darkgreen}{1235} & 522 & 660 \\
    \midrule
    \multicolumn{13}{c}{\textbf{Double Precision}} \\
    \midrule
	  $10^{-4}$ & 3479 & 1940 & 1324 & \boldcolor{darkgreen}{3396} & 1838 & 1325 & 4109 & 2162 & 1337 & 4262 & 2232 & 1359 \\
	  $10^{-6}$ & 10250 & 516 & 9273 & \boldcolor{darkgreen}{10157} & 466 & 9301  & 10206 & 491 & 9349 & 10778 & 1039 & 9310 \\
	  $10^{-8}$ & \boldcolor{darkgreen}{11087} & 1008 & 9336 & 11110 & 997 & 9263  & 11365 & 1103 & 9236 & 12189 & 1907 & 9236 \\
	  $10^{-10}$ & \boldcolor{darkgreen}{13067} & 2205 & 9327 & 13475 & 2361 & 9256 & 14470 & 3496 & 9247 & 16772 & 4495 & 9719 \\
    \bottomrule
  \end{tabular}
  }
\end{table}

\begin{table}[htbp]
  \centering
  \caption{Same as Table~\ref{tab:benchmarks-m1}, but for the AMD 3790X.}
  \label{tab:benchmarks-amd}
  {\scriptsize
  \begin{tabular}{l l l l l l l l l l l l l}
    \toprule
    & \multicolumn{3}{c}{\textbf{\blasmtl (1)}} & \multicolumn{3}{c}{\textbf{\blasmtl (5)}} & \multicolumn{3}{c}{\textbf{\blasmtl (10)}} & \multicolumn{3}{c}{\textbf{FFT-M2L}} \\
    $\bigO{\bm{\varepsilon}}$ & $\mathbsf{T}_{\text{FMM}}$ & $\mathbsf{T}_{\text{M2L}}$ & $\mathbsf{T}_{\text{P2P}}$ & $\mathbsf{T}_{\text{FMM}}$ & $\mathbsf{T}_{\text{M2L}}$ & $\mathbsf{T}_{\text{P2P}}$ & $\mathbsf{T}_{\text{FMM}}$ & $\mathbsf{T}_{\text{M2L}}$ & $\mathbsf{T}_{\text{P2P}}$ & $\mathbsf{T}_{\text{FMM}}$ & $\mathbsf{T}_{\text{M2L}}$ & $\mathbsf{T}_{\text{P2P}}$ \\
    \midrule
    \multicolumn{13}{c}{$\bm{N = 1 \times 10^6$}} \\
    \midrule
    \multicolumn{13}{c}{\textbf{Single Precision}} \\
    \midrule
    $10^{-3}$ & 126 & 16 & 110 & \boldcolor{darkgreen}{121} & 5 & 110 & \boldcolor{darkgreen}{121} & 6 & 110 & 122 & 11 & 110 \\
    $10^{-4}$ & \boldcolor{darkgreen}{130} & 18 & 110 & 131 & 12 & 110 & 137 & 19 & 110 & 149 & 38 & 110 \\
    \midrule
    \multicolumn{13}{c}{\textbf{Double Precision}} \\
    \midrule
    $10^{-6}$ & \boldcolor{darkgreen}{299} & 64 & 221 & 365 & 121 & 225 & 377 & 136 & 224 & 481 & 240 & 221 \\
    $10^{-8}$ & \boldcolor{darkgreen}{433} & 179 & 222 & 534 & 262 & 224 & 548 & 285 & 224 & 730 & 449 & 221 \\
    $10^{-10}$ & \boldcolor{darkgreen}{816} & 526 & 221 & 975 & 652 & 223 & 1008 & 707 & 223 & 1191 & 893 & 221 \\
    \midrule
    \multicolumn{13}{c}{\textbf{$\bm{N = 8 \times 10^6}$}} \\
    \midrule
    \multicolumn{13}{c}{\textbf{Single Precision}} \\
    \midrule
    $10^{-3}$ & 1110 & 139 & 932 & 1115 & 129 & 932 & 1104 & 136 & 936 & \boldcolor{darkgreen}{1077} & 130 & 931 \\
    $10^{-4}$ & \boldcolor{darkgreen}{1163} & 208 & 932 & 1219 & 211 & 933 & 1216 & 235 & 936 & 1209 & 255 & 931 \\
    \midrule
    \multicolumn{13}{c}{\textbf{Double Precision}} \\
    \midrule
    $10^{-6}$ & \boldcolor{darkgreen}{3261} & 1252 & 1890 & 3518 & 1466 & 1922 & 3732 & 1705 & 1921 & 3433 & 1415 & 1889 \\
    $10^{-8}$ & \boldcolor{darkgreen}{4739} & 2597 & 1892 & 5492 & 3228 & 1916 & 5514 & 3319 & 923 & 5440 & 3276 & 1889 \\
    $10^{-10}$ & \boldcolor{darkgreen}{8650} & 6343 & 1886 & 9918 & 7310 & 1917 & 9654 & 7140 & 1922 & 8903 & 6449 & 1886 \\
    \bottomrule
  \end{tabular}
  }
\end{table}

\begin{table}[htbp]
  \centering
  \caption{Throughput and required storage of data associated with the \mtl operator when using \blasmtl or \fftmtl for our benchmark problems on the M1 Pro. We observe that the throughput of the \blasmtl method in single precision is low until the associated data sizes are large enough to fully utilise available cache, at which point the fixed costs of setting up the calculation are amortized and we observe high throughputs which increase with accuracy and hence data sizes. We observe that the data associated with the \fftmtl approach is always larger than the \blasmtl, and as such we always operate at the bandwidth limit of the method - which saturates at approximately 120 GFLOP/s in single precision and 80 GFLOP/s in double precision for the M1 Pro.}
  \label{tab:throughput-m1}
  {\scriptsize
  \begin{tabular}{lcccc}
    \toprule
    $\bigO{\bm{\varepsilon}}$ & \multicolumn{2}{c}{\textbf{M2L Throughput (GFLOP/s)}} & \multicolumn{2}{c}{\textbf{M2L Storage (MB)}} \\
                              & \blasmtl & FFT-M2L & \blasmtl & FFT-M2L \\
    \midrule
    \multicolumn{5}{c}{$\bm{N = 1 \times 10^6$}} \\
    \midrule
    \multicolumn{5}{c}{\textbf{Single Precision}} \\
    \midrule
    $10^{-3}$   & 27 & 113   &  0.15 & 3.65   \\
    $10^{-4}$   & 100  & 101 &  0.79 & 8.12   \\
    \midrule
    \multicolumn{5}{c}{\textbf{Double Precision}} \\
    \midrule
    $10^{-6}$   & 53  & 67  & 2.32  &  51.19 \\
    $10^{-8}$   & 83  & 53  & 4.80  & 117.00 \\
    $10^{-10}$  & 194 & 83  & 33.91 & 223.44 \\
    \midrule
    \multicolumn{5}{c}{$\bm{N = 8 \times 10^6$}} \\
    \midrule
    \multicolumn{5}{c}{\textbf{Single Precision}} \\
    \midrule
    $10^{-3}$ & 38  & 110 & 0.22 & 3.65  \\
    \midrule
    \multicolumn{5}{c}{\textbf{Double Precision}} \\
    \midrule
    $10^{-4}$   & 30  & 59 & 0.55 & 16.25   \\
    $10^{-6}$   & 51  & 52 & 2.31  &  51.19 \\
    $10^{-8}$   & 115 & 68 & 9.47 & 117.00\\
    $10^{-10}$  & 191 & 58 & 33.90 & 223.44\\
    \bottomrule
  \end{tabular}
  }
\end{table}

\begin{table}[htbp]
  \centering
  \caption{Same as Table~\ref{tab:throughput-m1}, but for the AMD 3790X. Peak throughput when using \fftmtl saturates at approximately 80 GFLOP/s in single precision and 40 GFLOP/s in double precision which approximately matches the differences in memory throughputs on both architectures shown in Table \ref{tab:sec:appendix:hardware_and_software}.}
  \label{tab:throughput-amd}
  {\scriptsize
  \begin{tabular}{lcccc}
    \toprule
    $\bigO{\bm{\varepsilon}}$ & \multicolumn{2}{c}{\textbf{M2L Throughput (GFLOP/s)}} & \multicolumn{2}{c}{\textbf{M2L Storage (MB)}} \\
                              & \blasmtl & FFT-M2L & \blasmtl & FFT-M2L \\
    \midrule
    \multicolumn{5}{c}{$\bm{N = 1 \times 10^6$}} \\
    \midrule
    \multicolumn{5}{c}{\textbf{Single Precision}} \\
    \midrule
    $10^{-3}$   & 19  & 82  & 0.15  & 3.65   \\
    $10^{-4}$   & 88 & 53   & 0.78  & 8.13 \\
    \midrule
    \multicolumn{5}{c}{\textbf{Double Precision}} \\
    \midrule
    $10^{-6}$   & 85 & 28  & 4.38 & 51.19 \\
    $10^{-8}$   & 71 & 36  & 9.47 & 117.00 \\
    $10^{-10}$  & 77 & 36  & 33.91 & 223.44 \\
    \midrule
    \multicolumn{5}{c}{$\bm{N = 8 \times 10^6$}} \\
    \midrule
    \multicolumn{5}{c}{\textbf{Single Precision}} \\
    \midrule
    $10^{-3}$   & 28  & 55  & 0.22  & 3.65   \\
    $10^{-4}$   & 33 & 64   & 0.33   & 8.13   \\
    \midrule
    \multicolumn{5}{c}{\textbf{Double Precision}} \\
    \midrule
    $10^{-6}$   & 40 & 33  & 4.38 & 51.19 \\
    $10^{-8}$   & 44 & 36  & 9.47 & 117.00 \\
    $10^{-10}$  & 66 & 39  & 33.90 & 223.43 \\
    \bottomrule
  \end{tabular}
  }
\end{table}

\begin{table}
    \centering
    \scriptsize
    \caption{Optimal parameters for achieving an error of $\bigO{\varepsilon}$ calculated using equation \eqref{eq:l2_error} when using the \fftmtl for the Laplace \kifmm for benchmark problems with varying number of particles $N$ distributed in a unit box. $P_{\text{e}}$ is the expansion order used to construct the equivalent surface. `Batch Size' refers to the amount of data loaded into L1 cache per thread, as described in Appendix~\ref{appendix:fft_m2l}. `Total Setup Time' is the time taken to setup up all required buffers and index pointers, run precomputations, create the tree as well as maps required to lookup data required at runtime by the \mtl operation. We also separately list the times required for buffer and index pointer creation (`Buffers \& Pointers'), the creation of the maps required to lookup multipole data associated with the interaction list of each target box (`Map Creation', described in Appendix \ref{sec:fft_m2l_analysis}), the time required to create the octree, which includes a locational encoding for point data (`Tree'), and the time required to perform the relevant FFTs and data organization for the \mtl operation (`M2L Precalculation').}
    \label{tab:app:grid_search_laplace_fft_m2l}
\begin{minipage}{\textwidth}
    \subcaption{Apple M1 Pro (Single Precision, FFT M2L)}
    \centering
    \begin{tabular}{p{0.5cm}p{0.3cm}p{0.3cm}ccp{1cm}p{1cm}p{1cm}p{1cm}p{1cm}}
        \toprule
        $N$ & $\bigO{\bm{\varepsilon}}$ & $\mathbsf{P}_{\text{e}}$ & \textbf{Batch Size} & \textbf{Tree Depth} & \textbf{Tree (ms)} & \textbf{Buffers \& Pointers (ms)} & \textbf{Map Creation (ms)} & \textbf{M2L Precalculation (ms)} & \textbf{Total Setup (ms)}\\
        \midrule
        \multirow{2}{*}{$1 \times 10^6$}
            & $10^{-3}$ & 3 & 128 & 5 & 461 & 34 & 3 & 696 & 1200 \\
            & $10^{-4}$ & 4 & 64  & 5 & 461 & 40 & 3 & 707 & 1219 \\
        \midrule
        \multirow{1}{*}{$8 \times 10^6$}
            & $10^{-3}$ & 3 & 64  & 6 & 4064 & 289 & 51 & 800 & 5245 \\
        \bottomrule
    \end{tabular}
\end{minipage}

    \vspace{0.5em}
\begin{minipage}{\textwidth}
    \subcaption{Apple M1 Pro (Double Precision, FFT M2L)}
    \centering
    \begin{tabular}{p{0.5cm}p{0.3cm}p{0.3cm}ccp{1cm}p{1cm}p{1cm}p{1cm}p{1cm}}
        \toprule
        $N$ & $\bigO{\bm{\varepsilon}}$ & $\mathbsf{P}_{\text{e}}$ & \textbf{Batch Size} & \textbf{Tree Depth} & \textbf{Tree (ms)} & \textbf{Buffers \& Pointers (ms)} & \textbf{Map Creation (ms)} & \textbf{M2L Precalculation (ms)} & \textbf{Total Setup (ms)}\\
        \midrule
        \multirow{3}{*}{$1 \times 10^6$}
            & $10^{-6}$  & 6  & 128 & 5 & 448 & 97 & 4 & 841  & 1448 \\
            & $10^{-8}$  & 8  & 32  & 4 & 357 & 23 & 1 & 891  & 1484 \\
            & $10^{-10}$ & 10 & 64  & 4 & 376 & 48 & 1 & 1335 & 2435 \\
        \midrule
        \multirow{3}{*}{$8 \times 10^6$}
            & $10^{-4}$  & 4  & 128 & 6 & 3932 & 437 & 31 & 853  & 5301 \\
            & $10^{-6}$  & 6  & 16  & 5 & 3121 & 120 & 3 & 840  & 4181 \\
            & $10^{-8}$  & 8  & 64  & 5 & 3122 & 192 & 3 & 1191 & 4799 \\
            & $10^{-10}$ & 10 & 32  & 5 & 3166 & 375 & 4 & 1825 & 6239 \\
        \bottomrule
    \end{tabular}
\end{minipage}

    \vspace{1em}
\begin{minipage}{\textwidth}
    \subcaption{AMD 3790X (Single Precision, FFT M2L)}
    \centering
    \begin{tabular}{p{0.5cm}p{0.3cm}p{0.3cm}ccp{1cm}p{1cm}p{1cm}p{1cm}p{1cm}}
        \toprule
        $N$ & $\bigO{\bm{\varepsilon}}$ & $\mathbsf{P}_{\text{e}}$ & \textbf{Batch Size} & \textbf{Tree Depth} & \textbf{Tree (ms)} & \textbf{Buffers \& Pointers (ms)} & \textbf{Map Creation (ms)} & \textbf{M2L Precalculation (ms)} & \textbf{Total Setup (ms)}\\
        \midrule
        \multirow{2}{*}{$1 \times 10^6$}
            & $10^{-3}$ & 3 & 64 & 4 & 557  & 7  & 1 & 556 & 1138 \\
            & $10^{-4}$ & 4 & 32 & 4 & 534  & 9  & 1 & 586 & 1145 \\
        \midrule
        \multirow{2}{*}{$8 \times 10^6$}
            & $10^{-3}$ & 3 & 32 & 5 & 5253 & 59 & 7 & 738 & 6247 \\
            & $10^{-4}$ & 4 & 32 & 5 & 5262 & 87 & 7 & 777 & 6284 \\
        \bottomrule
    \end{tabular}
\end{minipage}

    \vspace{0.5em}
\begin{minipage}{\textwidth}
    \subcaption{AMD 3790X (Double Precision, FFT M2L)}
    \centering
    \begin{tabular}{p{0.5cm}p{0.3cm}p{0.3cm}ccp{1cm}p{1cm}p{1cm}p{1cm}p{1cm}}
        \toprule
        $N$ & $\bigO{\bm{\varepsilon}}$ & $\mathbsf{P}_{\text{e}}$ & \textbf{Batch Size} & \textbf{Tree Depth} & \textbf{Tree (ms)} & \textbf{Buffers \& Pointers (ms)} & \textbf{Map Creation (ms)} & \textbf{M2L Precalculation (ms)} & \textbf{Total Setup (ms)}\\
        \midrule
        \multirow{3}{*}{$1 \times 10^6$}
            & $10^{-6}$  & 6  & 64 & 4 & 478  & 17  & 1 & 823  & 1364 \\
            & $10^{-8}$  & 8  & 32 & 4 & 473  & 51  & 2 & 1124 & 1988 \\
            & $10^{-10}$ & 10 & 16 & 4 & 472  & 77  & 1 & 1632 & 3453 \\
        \midrule
        \multirow{3}{*}{$8 \times 10^6$}
            & $10^{-6}$  & 6  & 32 & 5 & 5335 & 258 & 7 & 1051 & 7390 \\
            & $10^{-8}$  & 8  & 32 & 5 & 5327 & 451 & 7 & 1463 & 8445 \\
            & $10^{-10}$ & 10 & 32 & 5 & 5323 & 672 & 7 & 2130 & 9943 \\
        \bottomrule
    \end{tabular}
\end{minipage}

\end{table}

\begin{table}[htbp]
    \centering
    \scriptsize
    \caption{Optimal parameters for achieving an error of $\bigO{\varepsilon}$ calculated using \eqref{eq:l2_error} when using \blasmtl for the Laplace \kifmm for our benchmark problems. $P_{\text{e}}$ is the expansion order for the equivalent surface, $P_{\text{c}}$ for the check surface. `Singular Value Threshold' is the threshold below which associated singular values and vectors of the \mtl matrices are filtered out. '\mtl Compression' describes the reduction in size of the \mtl operator matrices with respect to no thresholding being applied. 'Oversamples' refers to oversampling in the \rsvd applied to the \mtl operator matrices.  `Total Setup Time' is the time taken to setup up all required buffers and index pointers, run precomputations, create the tree as well as maps required to lookup data required at runtime by the \mtl operation. We also separately list the times required for buffer and index pointer creation (`Buffers \& Pointers'), the creation of the map in Step 1 of the Algorithm described in Section \ref{sec:sub:algorithm} (`Map Creation'), the time required to create the octree, which includes a locational encoding for point data (`Tree'), and the time taken to perform the matrix compression procedure described in Section \ref{sec:blas_m2l:sub:algorithm_setup} (`M2L Precalculation').}
    \label{tab:app:grid_search_laplace_blas_m2l}

\begin{minipage}{\textwidth}
    \subcaption{Apple M1 Pro (Single Precision, BLAS M2L)}
    \centering
    \begin{tabular}{p{0.5cm}p{0.3cm}p{0.3cm}p{0.3cm}p{1cm}p{1cm}p{1cm}p{1cm}p{1cm}p{0.5cm}p{0.5cm}p{1cm}p{1cm}p{1cm}}
        \toprule
        $N$ & $\bigO{\bm{\varepsilon}}$ & $\mathbsf{P}_{\text{e}}$ & $\mathbsf{P}_{\text{c}}$ & \textbf{Singular Value Threshold} & \textbf{M2L Compression} & \textbf{Oversamples} & \textbf{Tree Depth} & \textbf{Tree (ms)} & \textbf{Buffers \& Pointers (ms)} & \textbf{Map Creation (ms)} & \textbf{M2L Precalculation (ms)} & \textbf{Total Setup (ms)} \\
        \midrule
        \multirow{2}{*}{$1 \times 10^6$}
            & $10^{-3}$ & 3 & 3 & $10^{-3}$ & 32 \%   & 5  & 5 & 439  & 30 & 683 & 6 & 1164 \\
            & $10^{-4}$ & 3 & 4 & --        & --      & 10 & 5 & 424  & 35 & 716 & 28 & 1208 \\
        \midrule
        \multirow{1}{*}{$8 \times 10^6$}
            & $10^{-3}$ & 3 & 3 & -- & -- & 10 & 6  & 3884 & 273 & 6156 & 9 & 10357 \\
        \bottomrule
    \end{tabular}
\end{minipage}

    \vspace{0.5em}

\begin{minipage}{\textwidth}
    \subcaption{Apple M1 Pro (Double Precision, BLAS M2L)}
    \centering
    \begin{tabular}{p{0.5cm}p{0.3cm}p{0.3cm}p{0.3cm}p{1cm}p{1cm}p{1cm}p{1cm}p{1cm}p{0.5cm}p{0.5cm}p{1cm}p{1cm}p{1cm}}
        \toprule
        $N$ & $\bigO{\bm{\varepsilon}}$ & $\mathbsf{P}_{\text{e}}$ & $\mathbsf{P}_{\text{c}}$ & \textbf{Singular Value Threshold} & \textbf{M2L Compression} & \textbf{Oversamples} & \textbf{Tree Depth} & \textbf{Tree (ms)} & \textbf{Buffers \& Pointers (ms)} & \textbf{Map Creation (ms)} & \textbf{M2L Precalculation (ms)} & \textbf{Total Setup (ms)} \\
        \midrule
        \multirow{3}{*}{$1 \times 10^6$}
            & $10^{-6}$  & 5 & 6 & $10^{-3}$ & 84 \% & 5  & 5 & 465  & 86 & 689 & 326 & 1599 \\
            & $10^{-8}$  & 8 & 10 & $10^{-3}$ & 97 \% & 10 & 5 & 458  & 229 & 692 & 4992 & 6696 \\
            & $10^{-10}$ & 9 & 11 & $10^{-7}$ & 84 \% & 10 & 4 & 363  & 36 & 67 & 9634 & 10687 \\
        \midrule
        \multirow{3}{*}{$8 \times 10^6$}
            & $10^{-4}$  & 3 & 5 & $10^{-3}$ & 58 \%  & 10 & 6  & 4069  & 443 & 6164 & 94 & 10813 \\
            & $10^{-6}$  & 5 & 6 & $10^{-3}$ & 84 \%  & 20 & 5  & 3209  & 92  & 662 & 368 & 4398 \\
            & $10^{-8}$  & 7 & 8 & $10^{-5}$ & 82 \%  & 20 & 5  & 3174  & 174 & 664 & 1713 & 5929 \\
            & $10^{-10}$ & 9 & 11 & $10^{-7}$ & 84 \% & 20 & 5  & 3203  & 310 & 655 & 10064 & 14994 \\
        \bottomrule
    \end{tabular}
\end{minipage}

    \vspace{1em}
\begin{minipage}{\textwidth}
    \subcaption{AMD 3790X (Single Precision)}
    \centering
    \begin{tabular}{p{0.5cm}p{0.3cm}p{0.3cm}p{0.3cm}p{1cm}p{1cm}p{1cm}p{1cm}p{1cm}p{0.5cm}p{0.5cm}p{1cm}p{1cm}p{1cm}}
        \toprule
        $N$ & $\bigO{\bm{\varepsilon}}$ & $\mathbsf{P}_{\text{e}}$ & $\mathbsf{P}_{\text{c}}$ & \textbf{Singular Value Threshold} & \textbf{M2L Compression} & \textbf{Oversamples} & \textbf{Tree Depth} &\textbf{Tree (ms)} & \textbf{Buffers \& Pointers (ms)} & \textbf{Map Creation (ms)} & \textbf{M2L Precalculation (ms)} & \textbf{Total Setup (ms)} \\
        \midrule
        \multirow{2}{*}{$1 \times 10^6$}
            & $10^{-3}$ & 3 & 3 & $10^{-3}$ & 32 \% & 20 & 4 & 476  & 5 & 25 & 26 & 545 \\
            & $10^{-4}$ & 3 & 4 & --        & --    & 5  & 4 & 463  & 6 & 26 & 142 & 650 \\
        \midrule
        \multirow{2}{*}{$8 \times 10^6$}
            & $10^{-3}$ & 3 & 3 & --        & --    & 5  & 5 & 5211 & 57 & 142 & 21 & 5590 \\
            & $10^{-4}$ & 3 & 4 & $10^{-3}$ & 58 \% & 10 & 5 & 5102 & 53 & 142 & 158 & 5550 \\
        \bottomrule
    \end{tabular}
\end{minipage}

\begin{minipage}{\textwidth}
    \subcaption{AMD 3790X (Double Precision)}
    \centering
    \begin{tabular}{p{0.5cm}p{0.3cm}p{0.3cm}p{0.3cm}p{1cm}p{1cm}p{1cm}p{1cm}p{1cm}p{0.5cm}p{0.5cm}p{1cm}p{1cm}p{1cm}}
        \toprule
        $N$ & $\bigO{\bm{\varepsilon}}$ & $\mathbsf{P}_{\text{e}}$ & $\mathbsf{P}_{\text{c}}$ & \textbf{Singular Value Threshold} & \textbf{M2L Compression} & \textbf{Oversamples} & \textbf{Tree Depth} &\textbf{Tree (ms)} & \textbf{Buffers \& Pointers (ms)} & \textbf{Map Creation (ms)} & \textbf{M2L Precalculation (ms)} & \textbf{Total Setup (ms)} \\
        \midrule
        \multirow{3}{*}{$1 \times 10^6$}
            & $10^{-6}$  & 5 & 6 & $10^{-5}$ & 69 \% & 10 & 4 & 474 & 14  & 50 & 966   & 1528 \\
            & $10^{-8}$  & 7 & 8 & $10^{-5}$ & 82 \% & 10 & 4 & 474 & 27  & 44 & 5132  & 5916 \\
            & $10^{-10}$ & 9 & 11 & $10^{-7}$ & 84 \% & 10 & 4 & 471 & 50  & 44 & 28864 & 30477 \\
        \midrule
        \multirow{3}{*}{$8 \times 10^6$}
            & $10^{-6}$  & 5 & 6 & $10^{-5}$ & 69 \% & 10 & 5 & 4915 & 124 & 354 & 965   & 6628 \\
            & $10^{-8}$  & 7 & 8 & $10^{-5}$ & 82 \% & 10 & 5 & 4928 & 319 & 347 & 5175  & 11188 \\
            & $10^{-10}$ & 9 & 11 & $10^{-7}$ & 84 \% & 20 & 5 & 4909 & 623 & 352 & 30328 & 37704 \\
        \bottomrule
    \end{tabular}
\end{minipage}
\end{table}

\section{Discussion}\label{sec:discussion}

\subsection{Comparison With FFT-M2L}

The central plot of Figure \ref{fig:field_translations} shows the evaluation of the check potential at a target box from a source box in its interaction list. This can be interpreted as a convolution of the equivalent densities at each source box with a matrix of kernel evaluations between the source box's equivalent surface and the target box's check surface if the surfaces are discretized to the same degree. This convolution is accelerated using the \fft, which replaces the matrix-matrix product with an element-wise Hadamard product. However, this product has low operational intensity, since the number of memory accesses is proportional to the number of \flops. We note that the Fourier transforms of the kernel evaluation matrices can be precomputed and cached, which we describe in Appendix \ref{appendix:fft_m2l:algorithm}.

PVFMM addressed this challenge by reformulating the Hadamard product to achieve high computational throughput. As detailed in Appendix~\ref{appendix:fft_m2l}, the key idea being to increase the arithmetic intensity so that the working data fills L1 cache. The estimated arithmetic intensity is given by

\begin{flalign*}
    \frac{8 \cdot 26 \cdot 64 \cdot b}{2\cdot(64 \cdot 26  + 8b)} \> \> \text{FLOPs/Accesses}
\end{flalign*}

from equation \eqref{eq:chpt:appendix:sec:fft_m2l:arithmetic_intensity}, where \( b \) is the batch size of target boxes processed per thread. In comparison to our method based on matrix-matrix products for which operational intensity increases as a function of rank, the operational intensity of the \fftmtl approach is fixed due to its reliance on a Hadamard product.

Both approaches require memory-bound data organizations at runtime. Our \blasmtl method duplicates multipole data associated with each source box via their transfer vectors, while \fftmtl reorganizes data by frequency in Fourier space, and then applies an inverse operation. These operations are local to each target box and can be efficiently multithreaded.

A key advantage of \fftmtl is its relatively short precomputation time in comparison to \blasmtl in double precision, as seen in Tables \ref{tab:app:grid_search_laplace_fft_m2l} and \ref{tab:app:grid_search_laplace_blas_m2l}. In both methods the \mtl precomputations depend only on \fmm parameters - not on particle positions—allowing reuse across multiple right-hand sides. In \blasmtl these precomputations correspond to the compression of the \mtl matrices, and in \fftmtl they correspond to the computation of the \fft of the matrices of kernel evaluations between the source box equivalent surface and the target box check surface. As a result in applications of the \fmm in which equations of the form of \eqref{eq:sec:introduction:potential} are solved repeatedly for a static distribution of source and target points, such as in \acrshort{bem} solvers, this portion of the setup cost of the \fmm can be amortized over many applications of the \fmm.

At each timestep of a simulation involving dynamic particles one must recompute the locational encoding of the particle positions identifying their associated leaf boxes, reset the index pointers between boxes and associated particles and buffers containing their associated expansion and/or potential data, as well as clear all buffers for stored data. The most significant point of difference between \fftmtl and \blasmtl are the mappings required to identify required multipole data at runtime when processing each target box. For \blasmtl this mapping is between transfer vectors and associated source/target box pairs at each tree level, as described in Step~1 of the algorithm in Section~\ref{sec:sub:algorithm}. For \fftmtl this map creates index pointers between each target box and the required Fourier transforms of kernel evaluations and multipole data for sources in its interaction list. Further details of this are provided in Appendix~\ref{sec:fft_m2l_analysis}.

From Tables \ref{tab:app:grid_search_laplace_fft_m2l} and \ref{tab:app:grid_search_laplace_blas_m2l} we note that map creation time for \blasmtl grows with tree depth and becomes a significant portion of the setup time in for deeper trees. For \fftmtl, this cost remains negligible in comparison as we only create index pointers local to each target box, rather than a map between transfer vectors and all associated source/target box pairs at a given tree level. Conversely, the actual \mtl precomputation is cheaper in single precision and at low expansion orders in double precision for \blasmtl than \fftmtl using our \rsvd-based approach, but becomes more expensive at high orders in double precision. As a result, total setup times are broadly similar across both methods at the accuracies tested in our benchmark problems — except at high orders in double precision, where \fftmtl is faster on both of the architectures tested particularly on the AMD architecture using OpenBLAS for linear algebra.

These results suggest that \fftmtl is better suited for simulations of dynamic particles, especially when using deep trees, where the map creation time cannot be amortized. It also benefits from being \textit{exact} for translation-invariant kernels and requires no tuning for kernel-specific ranks, unlike \blasmtl. However, \fftmtl has drawbacks. Its operational intensity is limited, and its performance depends on complex data access patterns and explicit \acrshort{simd} implementations that must be rewritten for each \cpu \acrshort{isa}. In contrast, \blasmtl uses well-optimized \blas routines, which benefit from widespread hardware and software optimization. Finally, \fftmtl assumes that upward and downward surfaces are discretized to the same degree to permit a convolution interpretation. In contrast, the \mfs method is often more stable when using more check than equivalent points, as shown in~\cite{barnett2008stability}. This is demonstrated in the optimal parameters we identify in Tables \ref{tab:app:grid_search_laplace_fft_m2l} and \ref{tab:app:grid_search_laplace_blas_m2l} where we see that we are often able to use lower order equivalent surfaces when using \blasmtl with a greater number of check points to achieve similar errors to those achieved when using \fftmtl.

Our experiments imply architecture dependent performance considerations for the \kifmm, for which we identify the following four cases:

\begin{itemize}
    \item \textbf{High Bandwidth, High Core Count}: If an architecture exhibits high bandwidth and high core-counts, both our \ptp and \blasmtl implementations are favorable, and one can use shallow trees. Examples include the NVidia Grace \cpu Superchip, with a total of 144 ARM v9 cores, with a peak bandwidth of 1 TB/s.
    \item \textbf{Low Bandwidth, High Core Count}: If bandwidth is relatively low, with a high core-count such as the AMD 3790X tested, there is relatively little to discern the \blasmtl and \fftmtl at lower accuracies and especially in single precision. However, the high core-count lends itself to using shallow trees, and a greater reliance on the optimized \ptp operation.
    \item \textbf{High Bandwidth, Low Core Count}: If bandwidth is high, with a modest core-count such as the Apple M1 Pro tested, the \blasmtl consistently outperforms the \fftmtl due to its higher operational intensity which grows with matrix size. In this setting one must generally use deeper trees, to account for the relatively poorer performance of our \ptp implementation with fewer cores. Additionally, this type of architecture is well suited for considering multiple source density vectors at once, especially in single precision.
    \item \textbf{Low Bandwidth, Low Core Count}: In this case it is likely that the bandwidth constrained effects of the \fftmtl will not be noticeable, and the \blasmtl and \fftmtl can be used interchangeably, again necessitating deeper trees due to the lower number of cores.
\end{itemize}

The implications for extending our implementation to a heterogenous setting are that the best performance could be achieved with shallow trees exploiting the efficiency of the \blasmtl with the bulk of the potential evaluation conducted using the \ptp operation implemented on \gpus. These steps are data independent and can be called asynchronously.

\section{Conclusion}\label{sec:conclusion}

In this paper we have shown that with suitable blocking and careful use of randomized \acrshortpl{svd} a \blas based \kifmm can be competitive with the state-of-the art \fft based approach for problems described by the Laplace kernel, though precise runtime and setup performance differences between both methods is dependent on both architecture and the desired accuracy in the evaluated potentials.

The advantage of our approach is its simpler implementation and the portability offered by \acrshort{blas}. Our implementation naturally extends to \gpus via batched-\blas operations, although latency and memory transfers must be carefully tuned to achieve good performance. \blasmtl operators also naturally extend to treating many \fmm source density vectors at the same time, giving potential for additional cache reuse on architectures with favorable bandwidth properties such as the M1 Pro tested in this work.

The main trade-off, in addition to its relatively longer setup time for high-accuracy evaluations, are more sources of error in the evaluated potential due to the various parameters required by our compression scheme, as well as the sensitivity of the scheme to these parameters which must be tuned for each kernel. We note that the relatively longer precomputation time of \blasmtl in comparison to the \fftmtl for certain experimental settings makes it well suited to applications in which this is amortized with respect to other setup costs, or the \fmm is repeatedly applied to a static point distribution with changing right hand sides.

Looking ahead, as both \cpu and \gpu architectures continue to receive significant hardware and software investment in \blas optimizations, we expect that algorithm designs leveraging these primitives will become increasingly advantageous.

\FloatBarrier

\begin{acks}

We are grateful to Matthew Scroggs for his valuable assistance in software development during this research, as well as Dhairya Malhotra for informative conversations regarding PVFMM.

\end{acks}

\bibliographystyle{ACM-Reference-Format}
\bibliography{references}


\begin{thebibliography}{33}


\ifx \showCODEN    \undefined \def \showCODEN     #1{\unskip}     \fi
\ifx \showISBNx    \undefined \def \showISBNx     #1{\unskip}     \fi
\ifx \showISBNxiii \undefined \def \showISBNxiii  #1{\unskip}     \fi
\ifx \showISSN     \undefined \def \showISSN      #1{\unskip}     \fi
\ifx \showLCCN     \undefined \def \showLCCN      #1{\unskip}     \fi
\ifx \shownote     \undefined \def \shownote      #1{#1}          \fi
\ifx \showarticletitle \undefined \def \showarticletitle #1{#1}   \fi
\ifx \showURL      \undefined \def \showURL       {\relax}        \fi
\providecommand\bibfield[2]{#2}
\providecommand\bibinfo[2]{#2}
\providecommand\natexlab[1]{#1}
\providecommand\showeprint[2][]{arXiv:#2}

\bibitem[Ope(2024)]%
        {OpenBLAS}
 \bibinfo{year}{2024}\natexlab{}.
\newblock \bibinfo{booktitle}{\emph{OpenBLAS: An optimized BLAS library}}.
\newblock
\urldef\tempurl%
\url{https://www.openblas.net/}
\showURL{%
\tempurl}
\newblock
\shownote{Accessed: 2024-06-13}.


\bibitem[Agullo et~al\mbox{.}(2014)]%
        {agullo2014task}
\bibfield{author}{\bibinfo{person}{Emmanuel Agullo},
  \bibinfo{person}{B{\'e}renger Bramas}, \bibinfo{person}{Olivier Coulaud},
  \bibinfo{person}{Eric Darve}, \bibinfo{person}{Matthias Messner}, {and}
  \bibinfo{person}{Toru Takahashi}.} \bibinfo{year}{2014}\natexlab{}.
\newblock \showarticletitle{Task-based FMM for multicore architectures}.
\newblock \bibinfo{journal}{\emph{SIAM Journal on Scientific Computing}}
  \bibinfo{volume}{36}, \bibinfo{number}{1} (\bibinfo{year}{2014}),
  \bibinfo{pages}{C66--C93}.
\newblock


\bibitem[Ambikasaran et~al\mbox{.}(2014)]%
        {ambikasaran2014fast}
\bibfield{author}{\bibinfo{person}{Sivaram Ambikasaran},
  \bibinfo{person}{Michael O'Neil}, {and} \bibinfo{person}{Karan~Raj Singh}.}
  \bibinfo{year}{2014}\natexlab{}.
\newblock \showarticletitle{Fast symmetric factorization of hierarchical
  matrices with applications}.
\newblock \bibinfo{journal}{\emph{arXiv preprint arXiv:1405.0223}}
  (\bibinfo{year}{2014}).
\newblock


\bibitem[{Apple Incorporated}(2024)]%
        {AppleAccelerate}
\bibfield{author}{\bibinfo{person}{{Apple Incorporated}}.}
  \bibinfo{year}{2024}\natexlab{}.
\newblock \bibinfo{booktitle}{\emph{Accelerate Framework}}.
\newblock
\urldef\tempurl%
\url{https://developer.apple.com/documentation/accelerate}
\showURL{%
\tempurl}
\newblock
\shownote{Accessed: 2024-06-13}.


\bibitem[Barnett and Betcke(2008)]%
        {barnett2008stability}
\bibfield{author}{\bibinfo{person}{Alex~H Barnett} {and} \bibinfo{person}{Timo
  Betcke}.} \bibinfo{year}{2008}\natexlab{}.
\newblock \showarticletitle{Stability and convergence of the method of
  fundamental solutions for Helmholtz problems on analytic domains}.
\newblock \bibinfo{journal}{\emph{J. Comput. Phys.}} \bibinfo{volume}{227},
  \bibinfo{number}{14} (\bibinfo{year}{2008}), \bibinfo{pages}{7003--7026}.
\newblock


\bibitem[Betcke et~al\mbox{.}(2024)]%
        {bempp_green_kernels}
\bibfield{author}{\bibinfo{person}{Timo Betcke}, \bibinfo{person}{Matthew
  Scroggs}, {and} \bibinfo{person}{Srinath Kailasa}.}
  \bibinfo{year}{2024}\natexlab{}.
\newblock \bibinfo{title}{Green Kernels - A Rust library for the evaluation of
  Green's function kernels}.
\newblock \bibinfo{howpublished}{\url{https://github.com/bempp/green-kernels}}.
\newblock
\newblock
\shownote{Accessed: 2024-06-17}.


\bibitem[Blanchard et~al\mbox{.}(2015)]%
        {blanchard2015scalfmm}
\bibfield{author}{\bibinfo{person}{Pierre Blanchard},
  \bibinfo{person}{B{\'e}renger Bramas}, \bibinfo{person}{Olivier Coulaud},
  \bibinfo{person}{Eric Darve}, \bibinfo{person}{Laurent Dupuy},
  \bibinfo{person}{Arnaud Etcheverry}, {and} \bibinfo{person}{Guillaume
  Sylvand}.} \bibinfo{year}{2015}\natexlab{}.
\newblock \showarticletitle{ScalFMM: A generic parallel fast multipole
  library}. In \bibinfo{booktitle}{\emph{SIAM Conference on Computational
  Science and Engineering (SIAM CSE 2015)}}.
\newblock


\bibitem[bolbot(2018)]%
        {thingiverse_3253610}
\bibfield{author}{\bibinfo{person}{bolbot}.} \bibinfo{year}{2018}\natexlab{}.
\newblock \bibinfo{title}{Old Ship}.
\newblock \bibinfo{howpublished}{Thingiverse}.
\newblock
\urldef\tempurl%
\url{https://www.thingiverse.com/thing:3253610}
\showURL{%
\tempurl}
\newblock
\shownote{Accessed: 2024-07-10}.


\bibitem[Bramas(2020)]%
        {bramas2020tbfmm}
\bibfield{author}{\bibinfo{person}{B{\'e}renger Bramas}.}
  \bibinfo{year}{2020}\natexlab{}.
\newblock \showarticletitle{TBFMM: A C++ generic and parallel fast multipole
  method library}.
\newblock \bibinfo{journal}{\emph{Journal of Open Source Software}}
  \bibinfo{volume}{5}, \bibinfo{number}{56} (\bibinfo{year}{2020}),
  \bibinfo{pages}{2444}.
\newblock


\bibitem[Cabrera et~al\mbox{.}(2021)]%
        {cabrera2021toward}
\bibfield{author}{\bibinfo{person}{Anthony Cabrera}, \bibinfo{person}{Seth
  Hitefield}, \bibinfo{person}{Jungwon Kim}, \bibinfo{person}{Seyong Lee},
  \bibinfo{person}{Narasinga~Rao Miniskar}, {and} \bibinfo{person}{Jeffrey~S
  Vetter}.} \bibinfo{year}{2021}\natexlab{}.
\newblock \showarticletitle{Toward performance portable programming for
  heterogeneous systems on a chip: A case study with qualcomm snapdragon soc}.
  In \bibinfo{booktitle}{\emph{2021 IEEE High Performance Extreme Computing
  Conference (HPEC)}}. IEEE, \bibinfo{pages}{1--7}.
\newblock


\bibitem[Coulaud et~al\mbox{.}(2008)]%
        {coulaud2008high}
\bibfield{author}{\bibinfo{person}{Olivier Coulaud}, \bibinfo{person}{Pierre
  Fortin}, {and} \bibinfo{person}{Jean Roman}.}
  \bibinfo{year}{2008}\natexlab{}.
\newblock \showarticletitle{High performance BLAS formulation of the
  multipole-to-local operator in the fast multipole method}.
\newblock \bibinfo{journal}{\emph{J. Comput. Phys.}} \bibinfo{volume}{227},
  \bibinfo{number}{3} (\bibinfo{year}{2008}), \bibinfo{pages}{1836--1862}.
\newblock


\bibitem[Coulaud et~al\mbox{.}(2010)]%
        {coulaud2010high}
\bibfield{author}{\bibinfo{person}{Olivier Coulaud}, \bibinfo{person}{Pierre
  Fortin}, {and} \bibinfo{person}{Jean Roman}.}
  \bibinfo{year}{2010}\natexlab{}.
\newblock \showarticletitle{High performance BLAS formulation of the adaptive
  fast multipole method}.
\newblock \bibinfo{journal}{\emph{Mathematical and Computer Modelling}}
  \bibinfo{volume}{51}, \bibinfo{number}{3-4} (\bibinfo{year}{2010}),
  \bibinfo{pages}{177--188}.
\newblock


\bibitem[Dongarra et~al\mbox{.}(2017)]%
        {dongarra2017extreme}
\bibfield{author}{\bibinfo{person}{Jack Dongarra}, \bibinfo{person}{Stanimire
  Tomov}, \bibinfo{person}{Piotr Luszczek}, \bibinfo{person}{Jakub Kurzak},
  \bibinfo{person}{Mark Gates}, \bibinfo{person}{Ichitaro Yamazaki},
  \bibinfo{person}{Hartwig Anzt}, \bibinfo{person}{Azzam Haidar}, {and}
  \bibinfo{person}{Ahmad Abdelfattah}.} \bibinfo{year}{2017}\natexlab{}.
\newblock \showarticletitle{With extreme computing, the rules have changed}.
\newblock \bibinfo{journal}{\emph{Computing in Science \& Engineering}}
  \bibinfo{volume}{19}, \bibinfo{number}{3} (\bibinfo{year}{2017}),
  \bibinfo{pages}{52--62}.
\newblock


\bibitem[Fong and Darve(2009)]%
        {Fong2009}
\bibfield{author}{\bibinfo{person}{William Fong} {and} \bibinfo{person}{Eric
  Darve}.} \bibinfo{year}{2009}\natexlab{}.
\newblock \showarticletitle{The black-box fast multipole method}.
\newblock \bibinfo{journal}{\emph{J. Comput. Phys.}} \bibinfo{volume}{228},
  \bibinfo{number}{23} (\bibinfo{year}{2009}), \bibinfo{pages}{8712--8725}.
\newblock


\bibitem[Fujiwara(2000)]%
        {fujiwara2000fast}
\bibfield{author}{\bibinfo{person}{Hiroyuki Fujiwara}.}
  \bibinfo{year}{2000}\natexlab{}.
\newblock \showarticletitle{The fast multipole method for solving integral
  equations of three-dimensional topography and basin problems}.
\newblock \bibinfo{journal}{\emph{Geophysical Journal International}}
  \bibinfo{volume}{140}, \bibinfo{number}{1} (\bibinfo{year}{2000}),
  \bibinfo{pages}{198--210}.
\newblock


\bibitem[Gazzoni~Filho et~al\mbox{.}(2024)]%
        {gazzoni2024pqc}
\bibfield{author}{\bibinfo{person}{D{\'e}cio~Luiz Gazzoni~Filho},
  \bibinfo{person}{Guilherme Brand{\~a}o}, \bibinfo{person}{Gora Adj},
  \bibinfo{person}{Arwa Alblooshi}, \bibinfo{person}{Isaac~A
  Canales-Mart{\'\i}nez}, \bibinfo{person}{Jorge Ch{\'a}vez-Saab}, {and}
  \bibinfo{person}{Julio L{\'o}pez}.} \bibinfo{year}{2024}\natexlab{}.
\newblock \showarticletitle{PQC-AMX: Accelerating Saber and FrodoKEM on the
  Apple M1 and M3 SoCs}.
\newblock \bibinfo{journal}{\emph{Cryptology ePrint Archive}}
  (\bibinfo{year}{2024}).
\newblock


\bibitem[Greengard and Rokhlin(1987)]%
        {Greengard1987}
\bibfield{author}{\bibinfo{person}{Leslie Greengard} {and}
  \bibinfo{person}{Vladimir Rokhlin}.} \bibinfo{year}{1987}\natexlab{}.
\newblock \showarticletitle{A fast algorithm for particle simulations}.
\newblock \bibinfo{journal}{\emph{Journal of computational physics}}
  \bibinfo{volume}{73}, \bibinfo{number}{2} (\bibinfo{year}{1987}),
  \bibinfo{pages}{325--348}.
\newblock


\bibitem[Halko et~al\mbox{.}(2011)]%
        {halko2011finding}
\bibfield{author}{\bibinfo{person}{Nathan Halko}, \bibinfo{person}{Per-Gunnar
  Martinsson}, {and} \bibinfo{person}{Joel~A Tropp}.}
  \bibinfo{year}{2011}\natexlab{}.
\newblock \showarticletitle{Finding structure with randomness: Probabilistic
  algorithms for constructing approximate matrix decompositions}.
\newblock \bibinfo{journal}{\emph{SIAM review}} \bibinfo{volume}{53},
  \bibinfo{number}{2} (\bibinfo{year}{2011}), \bibinfo{pages}{217--288}.
\newblock


\bibitem[{Intel Corporation}(2023)]%
        {intel64-ia32-optimization}
\bibfield{author}{\bibinfo{person}{{Intel Corporation}}.}
  \bibinfo{year}{2023}\natexlab{}.
\newblock \bibinfo{booktitle}{\emph{Intel 64 and IA-32 Architectures
  Optimization Reference Manual, Volume 1}}.
\newblock
\urldef\tempurl%
\url{https://www.intel.com/content/www/us/en/content-details/671488/intel-64-and-ia-32-architectures-optimization-reference-manual-volume-1.html}
\showURL{%
\tempurl}
\newblock
\shownote{Version 049, Last updated: September 5, 2023}.


\bibitem[Li et~al\mbox{.}(2014)]%
        {li2014kalman}
\bibfield{author}{\bibinfo{person}{Judith~Yue Li}, \bibinfo{person}{Sivaram
  Ambikasaran}, \bibinfo{person}{Eric~F Darve}, {and} \bibinfo{person}{Peter~K
  Kitanidis}.} \bibinfo{year}{2014}\natexlab{}.
\newblock \showarticletitle{A Kalman filter powered by-matrices for
  quasi-continuous data assimilation problems}.
\newblock \bibinfo{journal}{\emph{Water Resources Research}}
  \bibinfo{volume}{50}, \bibinfo{number}{5} (\bibinfo{year}{2014}),
  \bibinfo{pages}{3734--3749}.
\newblock


\bibitem[Longva(2024)]%
        {paradis2024}
\bibfield{author}{\bibinfo{person}{Andreas Longva}.}
  \bibinfo{year}{2024}\natexlab{}.
\newblock \bibinfo{title}{paradis: Parallel processing with disjoint indices in
  Rust}.
\newblock \bibinfo{howpublished}{\url{https://github.com/Andlon/paradis}}.
\newblock


\bibitem[Malhotra(2017)]%
        {Malhotra2017FastIntegral}
\bibfield{author}{\bibinfo{person}{D. Malhotra}.}
  \bibinfo{year}{2017}\natexlab{}.
\newblock \emph{\bibinfo{title}{Fast Integral Equation Solver for Variable
  Coefficient Elliptic PDEs in Complex Geometries}}.
\newblock Doctoral dissertation. \bibinfo{school}{The University of Texas at
  Austin}.
\newblock
\newblock
\shownote{Available at. http://hdl.handle.net/2152/63349}.


\bibitem[Malhotra and Biros(2015)]%
        {Malhotra2015}
\bibfield{author}{\bibinfo{person}{Dhairya Malhotra} {and}
  \bibinfo{person}{George Biros}.} \bibinfo{year}{2015}\natexlab{}.
\newblock \showarticletitle{PVFMM: A parallel kernel independent FMM for
  particle and volume potentials}.
\newblock \bibinfo{journal}{\emph{Communications in Computational Physics}}
  \bibinfo{volume}{18}, \bibinfo{number}{3} (\bibinfo{year}{2015}),
  \bibinfo{pages}{808--830}.
\newblock


\bibitem[Messner et~al\mbox{.}(2012)]%
        {messner2012optimized}
\bibfield{author}{\bibinfo{person}{Matthias Messner},
  \bibinfo{person}{B{\'e}renger Bramas}, \bibinfo{person}{Olivier Coulaud},
  {and} \bibinfo{person}{Eric Darve}.} \bibinfo{year}{2012}\natexlab{}.
\newblock \showarticletitle{Optimized M2L kernels for the Chebyshev
  interpolation based fast multipole method}.
\newblock \bibinfo{journal}{\emph{arXiv preprint arXiv:1210.7292}}
  (\bibinfo{year}{2012}).
\newblock


\bibitem[Miles(1993)]%
        {miles1993compute}
\bibfield{author}{\bibinfo{person}{Douglas Miles}.}
  \bibinfo{year}{1993}\natexlab{}.
\newblock \showarticletitle{Compute intensity and the FFT}. In
  \bibinfo{booktitle}{\emph{Proceedings of the 1993 ACM/IEEE conference on
  Supercomputing}}. \bibinfo{pages}{676--684}.
\newblock


\bibitem[{NVIDIA Corporation}(2024)]%
        {NVIDIA_cuBLAS}
\bibfield{author}{\bibinfo{person}{{NVIDIA Corporation}}.}
  \bibinfo{year}{2024}\natexlab{}.
\newblock \bibinfo{title}{{cuBLAS Library Documentation}}.
\newblock
\urldef\tempurl%
\url{https://docs.nvidia.com/cuda/cublas/index.html}
\showURL{%
\tempurl}
\newblock
\shownote{Accessed: 2024-01-29}.


\bibitem[Rahimian et~al\mbox{.}(2010)]%
        {rahimian2010petascale}
\bibfield{author}{\bibinfo{person}{Abtin Rahimian}, \bibinfo{person}{Ilya
  Lashuk}, \bibinfo{person}{Shravan Veerapaneni}, \bibinfo{person}{Aparna
  Chandramowlishwaran}, \bibinfo{person}{Dhairya Malhotra},
  \bibinfo{person}{Logan Moon}, \bibinfo{person}{Rahul Sampath},
  \bibinfo{person}{Aashay Shringarpure}, \bibinfo{person}{Jeffrey Vetter},
  \bibinfo{person}{Richard Vuduc}, {et~al\mbox{.}}}
  \bibinfo{year}{2010}\natexlab{}.
\newblock \showarticletitle{Petascale direct numerical simulation of blood flow
  on 200k cores and heterogeneous architectures}.
\newblock  (\bibinfo{year}{2010}), \bibinfo{pages}{1--11}.
\newblock


\bibitem[Takahashi et~al\mbox{.}(2012)]%
        {takahashi2012optimizing}
\bibfield{author}{\bibinfo{person}{Toru Takahashi}, \bibinfo{person}{Cris
  Cecka}, \bibinfo{person}{William Fong}, {and} \bibinfo{person}{Eric Darve}.}
  \bibinfo{year}{2012}\natexlab{}.
\newblock \showarticletitle{Optimizing the multipole-to-local operator in the
  fast multipole method for graphical processing units}.
\newblock \bibinfo{journal}{\emph{Internat. J. Numer. Methods Engrg.}}
  \bibinfo{volume}{89}, \bibinfo{number}{1} (\bibinfo{year}{2012}),
  \bibinfo{pages}{105--133}.
\newblock


\bibitem[Van~Zee and Van De~Geijn(2015)]%
        {VanZee2015}
\bibfield{author}{\bibinfo{person}{Field~G Van~Zee} {and}
  \bibinfo{person}{Robert~A Van De~Geijn}.} \bibinfo{year}{2015}\natexlab{}.
\newblock \showarticletitle{BLIS: A framework for rapidly instantiating BLAS
  functionality}.
\newblock \bibinfo{journal}{\emph{ACM Transactions on Mathematical Software
  (TOMS)}} \bibinfo{volume}{41}, \bibinfo{number}{3} (\bibinfo{year}{2015}),
  \bibinfo{pages}{1--33}.
\newblock


\bibitem[Virtanen et~al\mbox{.}(2020)]%
        {virtanen2020scipy}
\bibfield{author}{\bibinfo{person}{Pauli Virtanen}, \bibinfo{person}{Ralf
  Gommers}, \bibinfo{person}{Travis~E Oliphant}, \bibinfo{person}{Matt
  Haberland}, \bibinfo{person}{Tyler Reddy}, \bibinfo{person}{David
  Cournapeau}, \bibinfo{person}{Evgeni Burovski}, \bibinfo{person}{Pearu
  Peterson}, \bibinfo{person}{Warren Weckesser}, \bibinfo{person}{Jonathan
  Bright}, {et~al\mbox{.}}} \bibinfo{year}{2020}\natexlab{}.
\newblock \showarticletitle{SciPy 1.0: fundamental algorithms for scientific
  computing in Python}.
\newblock \bibinfo{journal}{\emph{Nature methods}} \bibinfo{volume}{17},
  \bibinfo{number}{3} (\bibinfo{year}{2020}), \bibinfo{pages}{261--272}.
\newblock


\bibitem[Wang et~al\mbox{.}(2021)]%
        {wang2021exafmm}
\bibfield{author}{\bibinfo{person}{Tingyu Wang}, \bibinfo{person}{Rio Yokota},
  {and} \bibinfo{person}{Lorena~A Barba}.} \bibinfo{year}{2021}\natexlab{}.
\newblock \showarticletitle{ExaFMM: a high-performance fast multipole method
  library with C++ and Python interfaces}.
\newblock \bibinfo{journal}{\emph{Journal of Open Source Software}}
  \bibinfo{volume}{6}, \bibinfo{number}{61} (\bibinfo{year}{2021}),
  \bibinfo{pages}{3145}.
\newblock


\bibitem[Williams et~al\mbox{.}(2009)]%
        {williams2009roofline}
\bibfield{author}{\bibinfo{person}{Samuel Williams}, \bibinfo{person}{Andrew
  Waterman}, {and} \bibinfo{person}{David Patterson}.}
  \bibinfo{year}{2009}\natexlab{}.
\newblock \showarticletitle{Roofline: an insightful visual performance model
  for multicore architectures}.
\newblock \bibinfo{journal}{\emph{Commun. ACM}} \bibinfo{volume}{52},
  \bibinfo{number}{4} (\bibinfo{year}{2009}), \bibinfo{pages}{65--76}.
\newblock


\bibitem[Ying et~al\mbox{.}(2004)]%
        {Ying2004}
\bibfield{author}{\bibinfo{person}{Lexing Ying}, \bibinfo{person}{George
  Biros}, {and} \bibinfo{person}{Denis Zorin}.}
  \bibinfo{year}{2004}\natexlab{}.
\newblock \showarticletitle{A kernel-independent adaptive fast multipole
  algorithm in two and three dimensions}.
\newblock \bibinfo{journal}{\emph{J. Comput. Phys.}} \bibinfo{volume}{196},
  \bibinfo{number}{2} (\bibinfo{year}{2004}), \bibinfo{pages}{591--626}.
\newblock


\end{thebibliography}


\appendix

\section{The FFT M2L Algorithm}\label{appendix:fft_m2l}
\subsection{Algorithm}\label{appendix:fft_m2l:algorithm}

We review the method introduced in \cite{Malhotra2015} and re-implemented in \cite{wang2021exafmm} for maximizing the arithmetic intensity of the evaluation of check potentials (\ref{eq:check_potential_matvec}) using the \fft below. We use the case of a one dimensional problem for clarity.

\begin{figure}[htbp]
    \centering
    \includegraphics[width=0.9\textwidth]{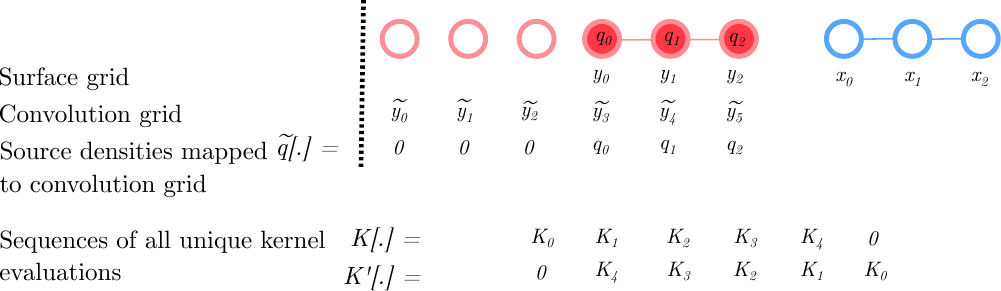}
    \caption{ We show a one dimensional M2L translation for $p=3$ expansions from the red source box $\{y_j\}_{j=0}^2$, embedded in a convolution grid $\{ \tilde{y}_j \}_{j=0}^{2P-1}$, with associated multipole expansion coefficients $\{q_j\}_{j=0}^2$ to a blue target box $\{x_i\}_{i=0}^2$. We associate the required sequences of kernel evaluations $K[.]$, and the flipped sequence $K'[.]$, and the sequence of source densities $\tilde{q}[.]$ associated with points on the convolution grid}
    \label{fig:sec:m2l:sub:formulation:sub:fft:one_dim_conv_grid}
\end{figure}

For an order $P$ multipole or local expansion, we described a check or equivalent surface as consisting of $P$ evenly spaced points along each axis as shown in Figure \ref{fig:p2m} in three dimensions. In one dimension, this would correspond to a line shown in Figure \ref{fig:sec:m2l:sub:formulation:sub:fft:one_dim_conv_grid}. We define the \textit{convolution grid} as an embedding of this surface into a grid defined by $\tilde{P} = 2P$ points along each axis through its volume, that encloses the grid describing the equivalent surface, which we call the \textit{surface grid}, and is aligned at a given corner of the surface grid. In three dimensions the convolution grid is instead a cube evenly discretized by $\tilde{P}$ points along each axis. We thus note that the convolution grid consists of $\tilde{P}^{\>d}$ points in dimension $d$.

We define a sequence of kernel evaluations as
\begin{equation}
    \nonumber
    K_j = K(x_c, \tilde{y}_j),
\end{equation}
where $x_c$ is a chosen point on the target check surface and $\tilde{y}_j$ are points on the convolution grid as shown in Figure \ref{fig:sec:m2l:sub:formulation:sub:fft:one_dim_conv_grid}. This sequence captures all the unique kernel evaluations between the points discretizing the source and target boxes. In the case of Figure \ref{fig:sec:m2l:sub:formulation:sub:fft:one_dim_conv_grid}, we choose $x_c = x_0$ and construct a sequence

\begin{equation}
    \nonumber
    K[j] = \begin{cases}
        K(x_c, \tilde{y}_{j+1}), & j = 0, ..., 2P-2 \\
        0, & j = 2P-1,
    \end{cases}
\end{equation}

where we use zero padding to handle the circular shift. We also define a sequence of densities on the convolution grid, defined through our embedding of the surface grid, placing zeros where densities from the surface grid are not mapped,

\begin{equation}
    \nonumber
    \tilde{q}[j] =
    \begin{cases}
        0, & j = 0,1...P-1, \\
        q[j-P], & j = P, ..., 2P-1,
    \end{cases}
\end{equation}

where $q[i]$, $i=0,...,P-1$, is the original sequence of densities on the surface grid.

We compute the check potential as a convolution of the flipped sequence $K'[2P-1 - i] = K[i]$ with the source densities placed on the convolution grid

\begin{flalign*}
    \nonumber
    \phi[i] = \sum_{j=0}^{2P-1}\tilde{q}[j]K'[(i-j-1)_{2P}]
\end{flalign*}

where $\phi[i]$ is the potential at $\phi(x_i)$.

Computed for a given box $\sigma$, finding the check potential consists of mapping the sequence of densities to the convolution grid corresponding to the multipole expansions for each box $A$ in its interaction list $I_{\sigma}$, computing the \dftfull of this sequence and computing the Hadamard product with the result with the \dft of the flipped sequence of kernel evaluations corresponding to that particular relative position between source and target box which can potentially be precomputed and cached. The \dft is accelerated with the \fft, however the component wise Hadamard product is of low operational intensity as each item of both sequence is used once per each required read and write operations.

\begin{figure}[h]
    \centering
    \includegraphics[width=0.6\textwidth]{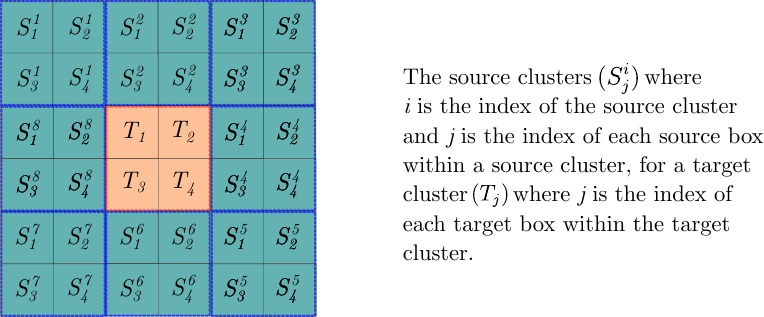}
    \caption{Source and target clusters illustrated in two dimensions. Here a target cluster consisting of four sibling quadrants is shown in ping, and the eight source clusters, which consist of the target cluster's parent's neighbours are shown in blue.}
    \label{fig:app:m2l_cluster}
\end{figure}

In three dimensions, this is an $O(\tilde{P}^{\>3})$ operation which requires $O(\tilde{P}^{\>3})$ memory accesses. For this case the authors of \cite{Malhotra2015} improve the arithmetic intensity by considering the interaction of eight source siblings with eight target siblings, allowing for efficient vectorisation. eight siblings together are referred to as a `cluster'. All the M2L translations for a target cluster will occur with boxes that are children of the neighbours of the cluster's shared parent, termed \textit{source clusters}. We illustrate this in two dimensions in Figure \ref{fig:app:m2l_cluster}. In two dimensions the source clusters form a halo consisting of eight clusters, around each target cluster. In three dimensions this halo consists of 26 source clusters. In two dimensions, there are 16 unique interaction pairs of source and target boxes between a given source cluster and a target cluster, correspondingly in three dimensions there are 64 such interactions. Thus each source and target cluster will have a 64 corresponding sequences of Fourier coefficients of kernel evaluations $K^{(i)}[]$ of length $\hat{P}=\tilde{P}^{\>3}$, where $i$ indexes the interaction between a source and target box contained in the source/target clusters being considered

\begin{align}
    &[K^{(1)}, K^{(2)}, \ldots, K^{(64)}] = \nonumber\\
    &\quad\big[ [K^{(1)}[0], K^{(1)}[1], \ldots, K^{(1)}[\hat{P}-1]], \nonumber\\
    &\quad\quad [K^{(2)}[0], K^{(2)}[1], \ldots, K^{(2)}[\hat{P}-1]], \nonumber\\
    &\quad\quad \ldots, \nonumber\\
    &\quad\quad [K^{(64)}[0], K^{(64)}[1], \ldots, K^{(64)}[\hat{P}-1]] \big], \nonumber
\end{align}

corresponding to 64 unique relative positions between the source and target boxes contained in these clusters. These sequences are permuted into \textit{frequency order},

\begin{equation}
    \begin{aligned}
        &\big[ \big[ K^{(1)}[0], K^{(2)}[0], \ldots, K^{(64)}[0] \big],\\
        &\quad \big[ K^{(1)}[1], K^{(2)}[1], \ldots, K^{(64)}[1] \big], \\
        &\quad \ldots, \\
        &\quad \big[ K^{(1)}[\hat{P}-1], K^{(2)}[\hat{P}-1], \ldots, K^{(64)}[\hat{P}-1] \big] \big]
    \end{aligned}
    \label{eq:sec:m2l:sub:formulation:sub:fft:permuted_fft_kernel}
\end{equation}

The Fourier coefficients $\hat{q}$ of the multipoles are ordered similarly. Consider the Fourier coefficients of a source cluster consisting of eight source boxes

\begin{flalign}
    &[\hat{q}^{(1)}, \hat{q}^{(2)}, ..., \hat{q}^{(8)}] = \\
    &\quad[[\hat{q}^{(1)}[0], ..., \hat{q}^{(1)}[\hat{P}-1]], \\
    &\quad[\hat{q}^{(2)}[0], ..., \hat{q}^{(2)}[\hat{P}-1]] \\
    &\quad\quad\dots\nonumber\\
    &\quad[\hat{q}^{(8)}[0], ..., \hat{q}^{(8)}[\hat{P}-1]]]
\end{flalign}

These are also permuted into frequency order,

\begin{flalign}
    &\quad[[\hat{q}^{(1)}[0], \hat{q}^{(2)}[0], ..., \hat{q}^{(8)}[0]], \\
    &\quad[\hat{q}^{(1)}[1], \hat{q}^{(2)}[1], ..., \hat{q}^{(8)}[1]] \\
    &\quad\quad\dots\nonumber\\
    &\quad[\hat{q}^{(1)}[\hat{P}-1], \hat{q}^{(2)}[\hat{P}-1], ..., \hat{q}^{(8)}[\hat{P}-1]] ]
\end{flalign}

The Hadamard product computation for the $k$'th frequency component of the check potential in Fourier space is then written as an $8 \times 8$ operation between all source boxes in an 8-cluster with the corresponding target boxes in an 8-cluster, where each element in a sequence is computed as,

\begin{equation}
    \hat{\phi}^{(i)}[k] =  K^{(i+8j)}[k] \cdot \hat{q}^{(i)}[k] + \hat{\phi}^{(i)}[k]
    \label{eq:chpt:appendix:sec:fft_m2l:hadamard_component}
\end{equation}

where $i,j \in [1, 8]$ and the sequence $\hat{\phi}^{(i)}[\cdot]$ corresponds to the check potential in Fourier space. The resulting sequence of check potentials in Fourier space is now in frequency order, and arranged by target cluster

\begin{flalign}
    &\quad[[\hat{\phi}^{(1)}[0], \hat{\phi}^{(2)}[0], ..., \hat{\phi}^{(8)}[0]], \\
    &\quad[\hat{\phi}^{(1)}[1], \hat{\phi}^{(2)}[1], ..., \hat{\phi}^{(8)}[1]] \\
    &\quad\quad\dots\nonumber\\
    &\quad[\hat{\phi}^{(1)}[\hat{P}-1], \hat{\phi}^{(2)}[\hat{P}-1], ..., \hat{\phi}^{(8)}[\hat{P}-1]] ]
\end{flalign}

These must be de-interleaved such that

\begin{flalign}
    &\quad[[\hat{\phi}^{(1)}[0], \hat{\phi}^{(1)}[1], ..., \hat{\phi}^{(1)}[\hat{P}-1]], \\
    &\quad[\hat{\phi}^{(2)}[0], \hat{\phi}^{(2)}[1], ..., \hat{\phi}^{(2)}[\hat{P}-1]] \\
    &\quad\quad\dots\nonumber\\
    &\quad[\hat{\phi}^{(8)}[0], \hat{\phi}^{(8)}[2], ..., \hat{\phi}^{(8)}[\hat{P}-1]] ]
\end{flalign}

for each target cluster, at which point an inverse \dft can be computed over the above sequences to recover the check potential.

The algorithm for problems in three dimensions for computing check potentials at each level $l \in [2, d]$ during the downward pass of the \kifmm consists of,

\begin{enumerate}
    \item In a precomputation step, compute sequences of kernel evaluations $K[\cdot]$ corresponding to all unique relative positions between a target cluster and source clusters in its halo. These are appropriately padded, and the \fft is computed for each one. The resulting sequences are then permuted into frequency order. Depending on the properties of the kernel, they can be stored once and scaled per level $l \in [2, d]$ in an octree, or must be stored for each level, $l \in [2, d]$, in an octree.
    \item First placing the multipole data for each source box on the convolution grid, compute the \fft of all multipole data at this level.
    \item Interleave the Fourier coefficients of the transformed multipole data for sets of sibling source boxes into frequency order.
    \item For each frequency, indexed by $k \in [0, \hat{P}-1]$, loop over all 26 source clusters in each target cluster's halo.
    For each source cluster, perform an $8 \times 8$ operation for the Hadamard product for the Fourier coefficients of the check potentials of each target cluster.
    Each element in the Hadamard product is computed with \eqref{eq:chpt:appendix:sec:fft_m2l:hadamard_component}.
    \item De-interleave the Fourier coefficients of the check potentials at each target cluster.
    \item Compute the inverse \fft to recover the check potentials for each target cluster.
\end{enumerate}

\subsection{Optimization}

Steps 2, 3 and 5 contain significant data organisation which must be done at runtime, and are therefore memory bound. However, as data associated with each box does not overlap these steps can be effectively parallelised with multithreading.

Additionally, Step 4 in the above algorithm can be effectively multithreaded over each frequency $k \in [0, \hat{P}-1]$. Each thread is then responsible for computing all the Hadamard products for all target clusters due to the source clusters in their halos at the $k$'th frequency. Threads are then pinned to cores for data locality.

In order to achieve high computational throughput the authors suggest that \textit{batches} of target clusters should be taken such that the L1 cache of each core is filled \cite{Malhotra2017FastIntegral}. With this, all source cluster directions are looped over for an entire batch for the $k$'th frequency components by each thread. For the architectures listed in Table \eqref{tab:sec:appendix:hardware_and_software} we find good performance with batch sizes of between 32 and 128 target clusters.

A drawback of this approach to \mtl involves computing \textit{all} translation directions for all source clusters relative to a target cluster, even if they do not appear in the interaction list of a target box in the target cluster. In this case the corresponding sequence of Fourier transformed kernel evaluations is replaced with zeros. This strategy therefore introduces some redundant computations, approximately $\sim10 \%$ according to the original authors \cite{Malhotra2017FastIntegral}, however this is worthwhile given the increased computational throughput.

Depending on the properties of the kernel, we can precompute and potentially scale the matrices corresponding to (\ref{eq:sec:m2l:sub:formulation:sub:fft:permuted_fft_kernel}) for all 26 relative positions between source and target clusters in 3 dimensions at each level $l$. Additionally, for sequences of kernel evaluations that correspond to real numbers, such as for (\ref{eq:sec:introduction:laplace_kernel}), the size of the resulting sequence of Fourier coefficients can be halved.

In order to further improve computational throughput, the original authors use explicit SIMD intrinsics for x86 architectures for the implementation of the $8\times8$ Hadamard product during the calculation in the parallel loop, as the sizes are too small to justify a \acrshort{blas} call and autovectorisers struggle to optimize the complex multiply add operations required on x86. In our implementation we follow the recommendations of the Intel architecture reference for this operation for AVX and AVX2 instruction sets \cite{intel64-ia32-optimization}. For Arm architectures, we use  NEON FCMA instructions, which contain special intrinsics for performing fused complex multiply and add operations. Our software also contains a generic autovectorised implementation, allowing our codes to run on common hardware targets supported by Rust's LLVM based compiler.

The frequency re-ordering together with reformulation as small efficient $8\times 8$ Hadamard products is key to make the method have high computational throughput. However, the permutations required to form and handle the frequency ordering results in practice in a complex code structure and care needs to be taken to do the re-ordering efficiently for it not to dominate execution time. In \cite{Malhotra2015} this was achieved by re-ordering mutable references to the actual data in a multithreaded loop. Access to mutable references by parallel threads is considered an anti-pattern in Rust, and despite being possible is not well supported due to the potential for race conditions and other parallel data access errors. Instead in our implementation we allocate new buffers to store the re-ordered data, which we can then iterate over in chunks corresponding to each frequency, and post-process the frequency ordered results for check potential back into Morton order. We note that there exists an emerging library for threadsafe parallel indirect access patterns in Rust, however we have not yet experimented with this \cite{paradis2024}.

\subsection{Analysis}\label{sec:fft_m2l_analysis}

As with our \blasmtl, the most challenging part of data access is ensuring the lookups required during the M2L for each box's interaction list, are contiguous. The multipole data in the halo of each target box is stored in a single contiguous buffer, in Morton order, over the source boxes at a given level. For each target box, we calculate its halo, and use a technique of index pointers to look up required data, whereby we store together the index position of the start of each multipole/local data for a given box within the global buffer storing all multipole/local data with its associated Morton key. Using index pointers, we create references to the multipole data of each target box's halo stored in frequency order. Therefore in the parallel loop we must lookup each reference in a loop over the halo data, and accumulate in a global buffer containing the check potentials for all target boxes, stored in frequency order and Morton order, un-permuting the check potentials from frequency order as a post-processing step. Similarly to our \blasmtl approach, the halo of each target box is calculated as a pre-processing step, and removes the need to explicitly construct and process interaction lists by indexing the octree at runtime. Efficient cache usage is achieved in this method by processing the halos of multiple target clusters at once for each frequency.

In our analysis we assume a uniformly refined tree such that the number of source boxes, $N_{\sigma}$, and the number of target boxes, $N_{\tau}$ at a given level $l \in [0, d]$ of an octree of depth $d$ are equal to $N_{\sigma} = N_{\tau} = 8^l$.

As Step 1 in the above algorithm can be done as a precomputation it does not enter our analysis for the runtime cost of this approach. In Step 2, we must place all multipole data for each source box on the convolution grid. In the \mfs in three dimensions the check and equivalent surfaces are discretized with

\begin{equation}
6 \cdot (P-1)^2 + 2
\label{eq:ncoeffs}
\end{equation}

points, this step therefore requires $\bigO{8^l \cdot P^{\>2}}$ read, addition and save operations, resulting in an operational intensity of,

\begin{flalign}
    \frac{1}{2} \> \> \text{FLOPs/Accesses}
\end{flalign}

The achieved operational intensity of the \fft in Step 2 is dependent both on the exact implementation of the operation to take advantage of \acrfull{ilp} optimizations, but also on the radix taken for the algorithm and the size of the data. From \cite{miles1993compute} operational intensity of this step can be estimated as

\begin{equation}
    \bigO{\log_2(P)}  \> \> \text{FLOPs/Accesses}
\end{equation}

Where we estimate the intensity for a radix of two in the \fft.

In Step 3, we must interleave the Fourier coefficients of the transformed multipole data for sets of sibling source boxes into frequency order. This involves $ 8^l \cdot \hat{P}$ read, addition and save operations, resulting in an operational intensity of

\begin{flalign}
    \frac{1}{2} \> \> \text{FLOPs/Accesses}
\end{flalign}

Step 4 involves the Hadamard product, naively computed this results in an operational intensity of $\bigO{1}$. In the scheme presented above, we are required to load $64 \cdot 26$ entries corresponding to the transformed kernel evaluations for the halo source boxes with respect to a target box from main memory. In addition to the loading of $8^l \cdot \hat{P}^{\>3}$ entries corresponding to the Fourier transformed multipole data in each source cluster, the Hadamard products result in $26 \cdot 64 \cdot 8^{l-1} \hat{P}^{\> 3}$ multiplication and addition operations over all target clusters at level $l$. These multiplications involve complex numbers, therefore each require a total of four multiplication and four addition operations to accumulate the result. Finally we require $8^l \cdot \hat{P}^{\>3}$ saves to store the resulting Fourier coefficients of check potentials. This results in an estimated operational intensity of,

\begin{equation}
    \frac{8 \cdot 26 \cdot 64 \cdot 8^{l-1} \hat{P}^{\> 3}}{2 \cdot (64 \cdot 26 + 2 \cdot 8^l \cdot \hat{P}^{\>3})} \> \> \text{FLOPs/Accesses}
\end{equation}

where the additional factor of two in the denominator comes from the fact that the data are complex numbers.

The \textit{arithmetic intensity}, which instead measures the ratio of \flops to memory traffic from \cpu cache rather than main memory is a function of L1 cache size in the above algorithm. In Step 4 we load a batch, $b$, of eight Fourier coefficients corresponding to each of $k \in [0, \hat{P}-1]$ frequencies into the L1 cache of each \cpu core, resulting in $64 \cdot 26$ loads for the Fourier coefficients of the kernel evaluations and $8b$ saves per \cpu core to save the check potentials data after Hadamard product. Each core then executes $26 \cdot 64 \cdot b$ multiplication and addition operations for each batch. This results in an arithmetic intensity of,

\begin{flalign}
    \label{eq:chpt:appendix:sec:fft_m2l:arithmetic_intensity}
    \frac{8 \cdot 26 \cdot 64 \cdot b}{2\cdot(64 \cdot 26  + 8b)} \> \> \text{FLOPs/Accesses}
\end{flalign}

per batch $b$ where again we notice that the data involved are complex numbers, assuming that the data of size $64 \cdot 26 + 8b$ entries, fits into the L1 cache of each core. We observe that while the operational intensity indicates that this step is memory bound, the observed throughput is high due to the fact that arithmetic intensity is an increasing function of L1 cache size. In the original presentation they report that up to 50 \% of theoretical peak performance is achieved on their tested 16 core Intel Xeon E5-2680 architectures with optimal choice of batch size $b=128$ with the additional use of explicit \acrshort{simd} and \fma instructions to perform the complex multiplications \cite{Malhotra2017FastIntegral}. On x86 architectures the complex multiplications involve shuffle operations which reduces the achieved throughput.

Step 5 performs a de-interleaving operation, similar to Step 2, and therefore also results in an operational intensity of

\begin{equation}
    \frac{1}{2} \> \> \text{FLOPs/Accesses}
\end{equation}

Similarly, the inverse \fft results in an operational intensity of

\begin{equation}
    \bigO{\log_2(P)}  \> \> \text{FLOPs/Accesses}
\end{equation}

\section{\blasmtl Algorithm Analysis}\label{app:analysis}

Consider the following matrix-matrix multiplication and accumulation operation for real matrices, such as \mtl matrices that arise from the Laplace kernel \eqref{eq:sec:introduction:laplace_kernel}

\begin{equation}
    \label{eq:matmul}
    \underset{M \times N}{\mathbsf{C}} =\underset{M \times K}{\mathbsf{A}} \> \> \times \> \> \underset{K \times N}{\mathbsf{B}} + \underset{M \times N}{\mathbsf{C}}
\end{equation}

where $\mathbsf{C} \in \mathbb{R}^{M \times N}$, $\mathbsf{A} \in \mathbb{R}^{M \times K}$, $\mathbsf{B} \in \mathbb{R}^{K \times N}$.

Each element in \eqref{eq:matmul} is calculated as $\mathbsf{C}_{ij} = \sum_{l=1}^K \mathbsf{A}_{il} \mathbsf{B}_{lj} + \mathbsf{C}_{ij}$. In total this requires $K$ multiply and $K-1$ addition operations per element in the sum and 1 more addition for the accumulation into $\mathbsf{C}$, resulting in $M \cdot N \cdot 2K$ \flops in total. The total number of accesses from memory involve reading $\mathbsf{A}$ and $\mathbsf{B}$ and reading and writing to $\mathbsf{C}$, resulting in $M \cdot K + K \cdot N + 2\cdot M \cdot N$ accesses.

This gives an estimate for the operational intensity of

\begin{equation}
    \label{eq:real_matmul_operational_intensity}
    \frac{2 \cdot M \cdot N \cdot K}{M \cdot K + K \cdot N + 2\cdot M \cdot N} \> \> \text{FLOPs}/\text{Accesses}
\end{equation}

Most modern \cpu and \gpu architectures support \fmafull instructions, which can effectively compute the multiply and add operations in a single instruction, with the cost of a single multiply operation. This leads to an approximate doubling of the throughput in the above estimates through an architecture, without increasing memory traffic.

In our analysis we assume a uniformly refined tree such that the number of source boxes, $N_{\sigma}$, and the number of target boxes, $N_{\tau}$ at a given level $l \in [0, d]$ of an octree of depth $d$ are equal to $N_{\sigma} = N_{\tau} = 8^l$.

As Step 1 can be performed as a precomputation it does not enter the runtime complexity estimate of the algorithm. In Step 2, we compute the compressed multipole expansions by applying $\mathbsf{S}^T$ to a buffer containing all the multipole expansions at a given level $l$, which are stored contiguously in our implementation. As these correspond to real matrix-matrix multiplications, this results in an operational intensity of,

\begin{equation*}
     \frac{2 \cdot 8^l \cdot k \cdot N_{\text{equiv}}}{2\cdot 8^l \cdot k + N_{\text{equiv}} \cdot k + 8^l \cdot N_{\text{equiv}} } \> \> \text{FLOPs/Accesses}
\end{equation*}

In Step 3, we are required to access and store the compressed multipole expansions associated with up to $N_i$ source boxes associated with each transfer vector as $i \in [1, 316]$. We can obtain an upper bound for this cost by assuming that \textit{every} source box at level $l$ appears once in each buffer, i.e. $N_i = N_{\sigma} = 8^l$ for all $i \in [1, 316]$.

\begin{figure}[h]
    \centering
    \includegraphics[width=0.3\textwidth]{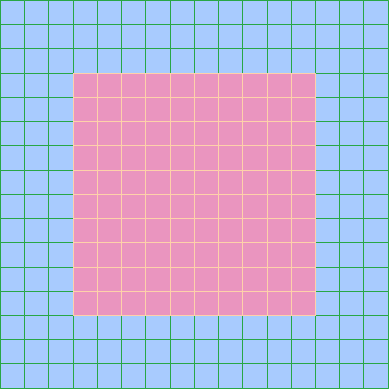}
    \caption{Interior boxes illustrated in pink in $\mathbb{R}^2$ for clarity, for a quadtree discretized to level $l=4$. Interior boxes increasingly dominate the total number of boxes with increasing tree level.}
    \label{fig:chpt::field_translation:sec:m2l:interior_boxes}
\end{figure}

This is true for all \textit{interior boxes} which we define as boxes which located at 3 box widths away from the boundary of the domain defined by the octree. We illustrate interior boxes in Figure \ref{fig:chpt::field_translation:sec:m2l:interior_boxes}. These boxes can be associated with any transfer vector $i \in [1, 316]$. With increasing tree level, $l$, interior boxes grow as $(2^l-6)^3$ and thus dominate the total number of boxes given by $8^l$.

With this assumption, we require $316 \cdot 8^l \cdot k$ read operations for all the compressed multipole expansions, $316 \cdot 8^l \cdot k$ addition operations for accumulating the multipole data and $316 \cdot 8^l \cdot k$ store operations to save the multipole data to the new buffers. This results in an operational intensity of

\begin{equation}
    \frac{1}{2} \> \> \text{FLOPs/Accesses}
\end{equation}

where we observe that this step is memory bound.

In Step 4 we compute the compressed check potentials over a loop of all the transfer vectors. For each transfer vector $i \in [1, 316]$, by the result \eqref{eq:real_matmul_operational_intensity} for real matrix-matrix multiplication, this results in an operational intensity of,

\begin{equation}
    \label{eq:app:operational_intensity_convolution_laplace}
     \frac{2 \cdot 8^l \cdot k \cdot k_t}{ 8^l \cdot k + k_t \cdot k + 2 \cdot  8^l \cdot k_t } + \frac{2 \cdot 8^l \cdot k \cdot k_t }{8^l \cdot k_t + k_t \cdot k + 2 \cdot 8^l \cdot k} \> \> \text{FLOPs/Accesses}
\end{equation}

where we've used the re-compressed form of $\mathbsf{C}_t$ with \eqref{eq:directional_compression}. This estimate is the worst case in which the intermediate result has to be accessed from main memory.

In Step 5, we need to accumulate compressed check potentials in buffers associated with each target box at level $l$. We can obtain an upper bound for this cost by assuming that the interaction lists of each target box is full, i.e. they are each $|I_{\tau}| = 189$. In which case we require $189 \cdot 8^l \cdot k$ read operations to lookup the compressed check potentials, $189 \cdot 8^l \cdot k$ addition operations for accumulating the compressed check potential data and $189 \cdot 8^l \cdot k$ store operations to save the compressed check potential data to the new buffers. This results in an operational intensity of

\begin{equation}
    \frac{1}{2} \> \> \text{FLOPs/Accesses}
\end{equation}

which we again observe is memory bound.

Finally, in Step 6, we calculate the uncompressed check potentials from the compressed form using a matrix-matrix multiplication. Again noticing that these correspond to real matrix-matrix products we see that this results in an operational intensity of,

\begin{equation}
    \frac{2 \cdot 8^l \cdot N_{\text{check}} \cdot k}{N_{\text{check}} \cdot k + 8^l \cdot k + 2 \cdot 8^l \cdot N_{\text{check}}} \> \> \text{FLOPs/Accesses}
\end{equation}

\section{Hardware \& Software}\label{app:optimal_parameters}

All Rust code was compiled using \texttt{opt-level=3}, \texttt{lto=true}, and \texttt{codegen-units=1} to maximize performance. CPU-specific optimizations were enabled via \texttt{-C target-cpu=native} and \texttt{-C target-feature} flags for each architecture (e.g., \texttt{AVX2} and \texttt{FMA} for \texttt{x86\_64}, \texttt{NEON} and \texttt{FMA} for \texttt{AArch64}).

\begin{table}[h]
    \caption{Hardware and software used in our benchmarks, for the Apple M1 Pro we report only the specifications of its `performance' CPU cores. We report per core cache sizes for L1/L2 and total cache size for L3. We note that the Apple M series of processors are designed with unusually large cache sizes, as well as unified memory architectures enabling rapid data access across specialized hardware units such as the performance CPU cores and the specialized matrix coprocessor used for BLAS operations when run with Apple's Accelerate framework \cite{AppleAccelerate}.}
    \label{tab:sec:appendix:hardware_and_software}
    \scriptsize
    \centering
    \begin{minipage}{\textwidth}
        \centering
        \begin{tabular}{l l l}
            \toprule
             & \textbf{Apple M1 Pro} & \textbf{AMD 3790X} \\
            \midrule
                Cache Line Size & 128 B & 64 B \\
                L1i/L1d & 192/128 KB & 32/32 KB \\
                L2      & 12 MB      & 512 KB \\
                L3      & 24 MB      & 134 MB \\
                Memory  & 16 GB      & 252 GB \\
                Max Clock Speed  & 3.2 GHz & 3.7 GHz \\
                Sockets/Cores/Threads & 1/8/8 & 1/32/64 \\
                Architecture & ArmV8.5 & x86\\
                SIMD Extensions & Neon & SSE, SSE2, AVX, AVX2 \\
                Memory Throughput & 200 GB/s & 102 GB/s \\
                \midrule
                BLAS & Apple Accelerate & Open BLAS \\
                LAPACK & Apple Accelerate & Open BLAS \\
                FFT & FFTW & FFTW \\
                Threading & Rayon & Rayon \\
            \bottomrule
            \end{tabular}
    \end{minipage}

\end{table}

\end{document}